\documentclass[%
 reprint,
superscriptaddress,
 amsmath,amssymb,
 aps, 
pra,
]{revtex4-1}

\usepackage{graphicx}
\usepackage{dcolumn}
\usepackage{bm}

\newcommand{\ii}{\mathrm{i}}
\newcommand{\tens}[1]{\overset{\text{\tiny$\leftrightarrow$}}{\mathbf{#1}}}

\usepackage{physics}
\usepackage{lipsum}
\usepackage{xcolor}
\usepackage[colorlinks=true,bookmarks=false,allcolors=blue]{hyperref}
\hypersetup{pdfstartview={FitH},pdfpagemode={UseNone}, breaklinks=true,            colorlinks,allcolors=blue, bookmarksopen=true, pdfnewwindow=true}
\usepackage[all]{hypcap}
\allowdisplaybreaks

\begin{document}

\preprint{APS/123-QED}

\title{Entangled Photon-pair Generation in Nonlinear Thin-films
}

\author{Elkin A. Santos}
\email{elkin.santos@uni-jena.de}
\affiliation{Institute of Applied Physics, Abbe Center of Photonics, Friedrich Schiller University Jena, Albert-Einstein-Str. 15, 07745 Jena,
Germany
}
\author{Maximilian A. Weissflog}
\affiliation{%
Institute of Applied Physics, Abbe Center of Photonics, Friedrich Schiller University Jena, Albert-Einstein-Str. 15, 07745 Jena,
Germany
}%
\affiliation{%
Max Planck School of Photonics, Hans-Kn\"oll-Straße 1, 07745 Jena, Germany
}%
\author{Thomas Pertsch}
\affiliation{%
Institute of Applied Physics, Abbe Center of Photonics, Friedrich Schiller University Jena, Albert-Einstein-Str. 15, 07745 Jena,
Germany
}%
\affiliation{Fraunhofer Institute for Applied Optics and Precision Engineering IOF, Albert-Einstein-Str. 7, 07745 Jena, Germany}
\author{Frank Setzpfandt}
\affiliation{%
Institute of Applied Physics, Abbe Center of Photonics, Friedrich Schiller University Jena, Albert-Einstein-Str. 15, 07745 Jena,
Germany
}%
\affiliation{Fraunhofer Institute for Applied Optics and Precision Engineering IOF, Albert-Einstein-Str. 7, 07745 Jena, Germany}
\author{Sina Saravi}%
\affiliation{Institute of Applied Physics, Abbe Center of Photonics, Friedrich Schiller University Jena, Albert-Einstein-Str. 15, 07745 Jena,
Germany
}


\date{\today}

\begin{abstract}
We develop a fully vectorial and non-paraxial formalism to describe spontaneous parametric down-conversion in nonlinear thin films. The formalism is capable of treating slabs with a sub-wavelength thickness, describe the associated Fabry-Pérot effects, and even treat absorptive nonlinear materials. With this formalism, we perform an in-depth study of the dynamics of entangled photon-pair generation in nonlinear thin films, to provide a needed theoretical understanding for such systems that have recently attracted much experimental attention as sources of photon pairs. As an important example, we study the far-field radiation properties of photon pairs generated from a high-refractive-index nonlinear thin-film with Zinc-Blende structure, that is deposited on a linear low-refractive-index substrate. In particular, we study the thickness-dependent effect of Fabry-Pérot interferences on the far-field radiation pattern of the photon pairs. We also pay special attention to study of entanglement generation, and find the conditions under which maximally polarization-entangled photon pairs can be generated and detected in such nonlinear thin-films.
\end{abstract}

\maketitle

\section{Introduction} 

The optical process of spontaneous parametric down-conversion (SPDC) in $\chi^{(2)}$-materials \cite{spdc}, in which a pump ($p$) photon of wavelength $\lambda_p$ can split into a pair of entangled signal ($s$) and idler ($i$) photons of lower energy, is an instrumental tool in optical quantum technologies. Predominant applications are quantum communication \cite{Flamini2019}, quantum cryptography \cite{Bennet,pirandola2020}, quantum imaging \cite{FrankRev,moreau} and sensing \cite{clark}, and testing fundamental quantum effects \cite{Horodecki}. The broad applicability of SPDC is due to its efficiency and versatility in generating pairs of entangled photons with high control over various optical degrees of freedom. 

Nonlinear systems involving SPDC constitute to date the dominant approach to generate entangled photon-pair states 
and promise tangible real-world applications in the near future \cite{wang2021}. Entangled photon pairs have been generated in single nonlinear bulk crystals \cite{Kwiat1995}, crossed bulk crystals \cite{Kwiat1999}, or bulk crystals in a Sagnac loop \cite{Hentschel}. More recently, several experiments of
photon-pair generation in much smaller nonlinear systems 
such as single GaAs nanowires \cite{saerens2023}, dielectric nanoantennas \cite{Marino2019}, and metasurfaces with subwavelength thicknesses \cite{ ,santiago2021,Santiago2022,sukho2022,Son2023} have been demonstrated. In particular, there has been a recent growing interest in pair generation in sub-wavelength-thin nonlinear crystals like lithium niobate and “Zinc-Blende” structures like gallium arsenide (GaAs) or gallium phosphide (GaP) \cite{Okoth2019,Chek_2021,Chek_2022}, as well as in van der Waals \cite{guo2023} or transition metal dichalcogenide (TMD) crystals \cite{weissflog2023tunable, trovatelloQuasiphasematchedDownconversionPeriodically2023}. 
Such nonlinear thin sources are interesting because they are not restricted by the longitudinal phase-matching condition, hence they are capable of generating photon pairs in a very wide spectral and angular range. This also motivates their use in quantum imaging applications, for pushing the resolution of such schemes to the limit \cite{vega2022,santos2022}.

However, to the best of our knowledge, there has not yet been a comprehensive
theoretical analysis of photon-pair generation in a thin and unstructured nonlinear slab, where the dynamics of entanglement generation is investigated and the conditions for generating a maximally entangled photon-pair state are identified.
Such an analysis, which would be facilitated by a general theoretical framework, is an essential missing link for further development of extremely thin and broadband sources of entangled photon pairs. This theoretical framework should incorporate the Fabry–Pérot-type interferences due to multiple reflections of signal, idler and also pump photons within the nonlinear slab. 
Moreover, such a theory should be able to treat absorptive nonlinear layers, particularly in light of the more prevalent use of TMD systems which provide an enormous enhancement of the nonlinear coefficients near their excitonic lines that at the same time can also cause enhanced material absorption in the system \cite{Krasnok}. Importantly, the formalism should also be able to describe the directional, spectral, and polarization properties of the generated photon pairs, including emission angles up to 90 degrees.

In this work, we develop such a theoretical framework for describing SPDC in nonlinear thin-films, and use it to study the far-field properties of the generated pairs. Our formalism is based on the Green’s function (GF) quantization method \cite{poddubny2016}. In this method, SPDC is described using a quantization scheme of the electric field operator that utilizes the classical GF of the system and local bosonic excitations in the medium-assisted field \cite{welsch1996,welsch1998}. The GF, as a classical quantity, contains all the linear properties of the system. In our work, we construct this classical GF using a developed theory for the GF of multilayered optical systems \cite{Sipe}. Due to the generality of the GF method, we develop a comprehensive, fully vectorial framework capable of treating any thickness of the slab, ultra-thin or thick. It can further be used to incorporate lossy layers, near- and far-field radiation in the non-paraxial regime, as well as the generation of guided photonic modes inside the source. In the current work, we focus on far-field properties of the photons that are generated outside the slab. Furthermore, our model allows to keep track of any polarization and directionality effects in the pair-generation process, which we will exploit in this paper to reconstruct the polarization states of entangled photon pairs. 

In this work we are treating SPDC in the low-gain regime of photon-pair generation. Yet recently, the application of the GF quantization method has also been extended to treat the high-gain regime of the interaction \cite{PhysRevResearch.5.043228}, where the same classical GF can be used for description of high-gain effects. This further emphasizes the versatility of the GF approach in description of parametric down-conversion under different conditions. We also note, that there are other methods for description of SPDC \cite{spdc,howell,Check2020,pevrina2011, pevrina2014, kitaeva}, some of which were also developed for description of low-gain SPDC in multilayer systems \cite{pevrina2011, pevrina2014, kitaeva}. Nontheless, the GF method provides a general yet unified formulation for treatment of such systems, which we employ to perform a detailed investigation of SPDC and identify entanglement properties in thin-films. Specially, the GF method can naturally include the effect of internal losses, and also it provides a way towards description of high-gain effects \cite{PhysRevResearch.5.043228}, which allows for study of thin-film systems in more advanced scenarios beyond this work.


In this investigation, we will utilize numerical calculations to explore the far-field characteristics of photon-pair generation via SPDC within a GaAs thin film. This is part of a larger category of nonlinear materials characterized by a Zinc-Blende structure, a type of nonlinear material with a cubic lattice and point group $\bar{4}3$m. This crystal structure is shared by various other prominent nonlinear materials, including III–V semiconductors such as gallium phosphide (GaP) and indium phosphide (InP), as well as II–VI semiconductors like zinc telluride (ZnTe), zinc selenide (ZnSe), and zinc sulfide (ZnS). Due to the cross-polarized nature of their nonlinear tensor, these materials are naturally predisposed for generating polarization-entangled Bell states \cite{weissflog2024,Chek_2022}. This holds significant promise for advancing various quantum information processing applications, ranging from quantum cryptography to quantum teleportation. By investigating the unique properties of photon-pair generation in GaAs and related nonlinear materials, we aim to contribute to the broader understanding and potential utilization of entangled photon pairs in cutting-edge quantum technologies.

The manuscript is organized as follows. In Sec.~(\ref{sec:JSP}) we introduce a general framework to calculate the far-field joint detection probability of photon pairs generated in a nonlinear slab in a layered geometry using the GF formalism, and compute the far-field joint probability distribution. In Sec.~(\ref{sec:GFs}), we explicitly write the general expression of the GF's of a dielectric slab that contains multiple reflections of the signal and idler fields. In Sec. (\ref{sec:pump}), the classical pump beam is treated. In Sec.~(\ref{sec:farfield}), we numerically compute the far-field radiation for a nonlinear source made of GaAs deposited on a silicon dioxide (SiO$_2$) substrate, and study its far-field properties. In Sec.~(\ref{sec:Qtomo}), we use a quantum tomography method to extract the polarization state of the generated photon pairs and show that one can generate maximally polarization-entangled photon pairs in such a simple system. Lastly, we summarize and conclude in Sec.~(\ref{sec:conc}).

\section{Far-Field Joint Spatial Probability in Fourier Domain}
\label{sec:JSP}

\begin{figure}[ht] 
\centering
\includegraphics[scale=1]{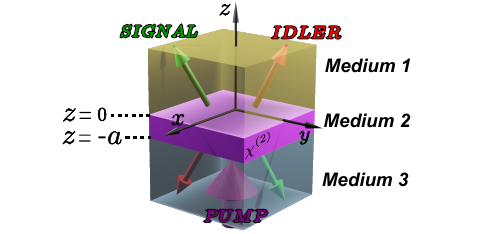}
\caption{Multilayered geometry where medium 2 is a nonlinear medium, in which SPDC happens due to a pump field impinging from medium 3.  The joint signal and idler field can radiate to outside the slab to medium 1 and medium 3, after experiencing multiple reflections inside medium 2.}
\label{fig:slab}
\end{figure}

Consider a general nonlinear thin-film photon-pair source, as depicted in Fig. \ref{fig:slab}. It comprises three layers with relative permittivities $\epsilon_1$, $\epsilon_2$, and $\epsilon_3$, where the nonlinear material (of thickness $a$) constitutes medium 2. Typically, medium 3 is a linear substrate and medium 1 is a linear cladding covering the nonlinear material (often air). Here, the system is illuminated by a pump beam from medium 3 propagating along the positive $z$-direction, and the down-converted photon pairs can be detected in media 1 or 3. It is important to note that the generated photon-pair amplitudes undergo multiple reflections inside the nonlinear slab. Although this paper does not delve into the dynamics of photon pairs that could be generated in the guided slab modes of medium 2, our formalism can handle such scenarios too. This is left to future works. In this work, we focus on studying the dynamics of photons generated into the free propagating modes of medium 1.

Using the GF method, we analyze SPDC in a nonlinear structure excited by a single-frequency pump beam $\mathbf{E}_p(\mathbf{r},t)=\mathbf{E}_p(\mathbf{r})\exp{-\ii\omega_pt}+ \text{c.c.}$. The far-field coincidence detection rate for a pair of signal and idler photons of different frequencies as a function of the spatial coordinates follows \cite{poddubny2016,weissflog2024}: 
\begin{align}\label{eq:prob_SI}
    \nonumber&\mathcal{R}(\mathbf{r}_s,\mathbf{r}_i, \mathbf{e}_s, \mathbf{e}_i,\omega_s,\omega_p-\omega_s)=\frac{8}{\pi} n_in_sr_i^2r_s^2\frac{(\omega_p-\omega_s)^3\omega_s^3}{c^6}\\ \nonumber
    &\times \bigg|\sum_{\alpha,\beta,\gamma}\sum_{\sigma_s,\sigma_i}e_{s,\sigma_s}e_{i,\sigma_i}\,\int_V\dd \mathbf{r}\,\chi^{(2)}_{\alpha\beta\gamma}(\mathbf{r})\\ 
    &\times E_{p,\gamma}(\mathbf{r})\,G_{\sigma_s\alpha}(\mathbf{r}_s,\mathbf{r},\omega_s)\,G_{\sigma_i\beta}(\mathbf{r}_i,\mathbf{r},\omega_p-\omega_s)\bigg|^2\,.
\end{align}

The detection rate, $\mathcal{R}\equiv\frac{\dd^4N}{\dd t\dd\Omega_s\dd\Omega_i\dd\omega_s} $, is the number of photon pairs given per units of solid angles $\dd\Omega_i$
(for the idler photon) and $\dd\Omega_s$ (for the signal photon), and
per unit of signal-photon angular frequency $\omega_s$. The idler angular
frequency is fixed by conservation of energy to $\omega_i=\omega_p-\omega_s$. The detection rate is also a function of the idler and signal photons' detection polarization, which are along the unitary $\mathbf{e}_s$ and $\mathbf{e}_i$
directions. The $\alpha$, $\beta$, $\gamma$, $\sigma_s$ and $\sigma_i$  indices run over the $x$, $y$, and $z$ cartesian coordinates. $G_{ij}(\mathbf{r,r'},\omega)$ are the tensor-components of the electric GF, $\chi^{(2)}_{\alpha\beta\gamma}(\mathbf{r})$ are the components of the second-order nonlinear tensor and the integral runs over the volume of nonlinearity $V$ characterized by the position vector $\mathbf{r}$. $\mathbf{r}_i$ and $\mathbf{r}_s$ are the positions for the detection of the idler and signal photons in the far-field, respectively, with $n_i$ and $n_s$ being the refractive indices of the detection medium at the idler and signal frequencies.

In our system, one can take advantage of the homogeneity in the $x$ and $y$ direction to work in the spatial frequencies domain. This approach offers an intuitive understanding of the far-field detection in terms of the angular modes of the signal and idler photons, simplifying numerical calculations. To do this,  the GF, which is a solution of $[\grad\times\grad\times-\frac{\omega^2}{c^2}\epsilon(\mathbf{r},\omega)]\tens{G}(\mathbf{r},\mathbf{r}',\omega)=\tens{I}\delta(\mathbf{r}-\mathbf{r}')$, with $\epsilon(\mathbf{r},\omega)$ being the relative permittivity of the system, can be represented as an inverse Fourier transform of the form \cite{lau2007},
\begin{equation}\label{eq:GF_fourier}
    \tens{G}(\mathbf{r},\mathbf{r}',\omega)=\frac{1}{(2\pi)^2}\int_{-\infty}^\infty\dd \mathbf{q}\;\tens{g}(\mathbf{q},z,z',\omega)\,e^{\ii\mathbf{q}\cdot(\mathbf{r}_\perp-\mathbf{r}'_\perp)}\,,
\end{equation}
where $\mathbf{q}=k_x\hat{x}+k_y\hat{y}$ and $\mathbf{r}_\perp=x\hat{x}+
y\hat{y}$ are the two-dimensional real wave-vector and spatial vector in the $x$-$y$
plane, respectively. We define $\int_{-\infty}^\infty\dd \mathbf{q}\equiv\int_{-\infty}^\infty\dd k_x\int_{-\infty}^\infty\dd k_y$. Similarly, for a classical pump field treated in the undepleted pump approximation we have,
\begin{equation}\label{eq:E_p_fourier} 
    \mathbf{E}_p(\mathbf{r})_=\frac{1}{(2\pi)^2}\int_{-\infty}^\infty\dd\mathbf{q}\;\mathbf{E}_p(\mathbf{q},z)\;e^{\ii \mathbf{q}\cdot\mathbf{r}_\perp}\,.
\end{equation}
Then, by introducing Eqs. (\ref{eq:GF_fourier}) and (\ref{eq:E_p_fourier}) into (\ref{eq:prob_SI}), we can separate the spatial integral into $\dd \mathbf{r}\rightarrow\dd z\,\dd\mathbf{r}_\perp$. After rearranging terms, we can solve for $\dd\mathbf{r}_\perp$ the integral
\begin{equation}
    \int_{-\infty}^{\infty}\dd \mathbf{r}_\perp\;e^{\ii(\mathbf{q}_p-\mathbf{q}_s-\mathbf{q}_i)\cdot \mathbf{r}_\perp}=(2\pi)^2\delta(\mathbf{q}_p-\mathbf{q}_s-\mathbf{q}_i)\,,
\end{equation}
which results into the joint detection rate
\begin{align}\label{eq:prob_SI_fourier2}
    \nonumber&\mathcal{R}(\mathbf{r}_s,\mathbf{r}_i, \mathbf{e}_s, \mathbf{e}_i,\omega_s,\omega_p-\omega_s)=\frac{8}{\pi} n_in_sr_i^2r_s^2\frac{(\omega_p-\omega_s)^3\omega_s^3}{c^6}\\&\times\bigg|\frac{1}{(2\pi)^4}\int_{-\infty}^\infty\int_{-\infty}^\infty \dd \mathbf{q}_s\,\dd \mathbf{q}_i\;\Tilde{\mathcal{R}}(\mathbf{q}_s,\mathbf{q}_i, z_s,z_i,\omega_s,\omega_p-\omega_s)\;\\\nonumber
    &\times e^{\ii(\mathbf{q}_s\cdot\mathbf{r}_{s\perp}+\mathbf{q}_i\cdot\mathbf{r}_{i\perp})}\bigg|^2\,,
\end{align}
with
\begin{align}\label{eq:JSA_SI}
    \nonumber &\Tilde{\mathcal{R}}(\mathbf{q}_s,\mathbf{q}_i, z_s,z_i,\omega_s,\omega_p-\omega_s)=   \sum_{\alpha,\beta,\gamma}\sum_{\sigma_s,\sigma_i}e_{s,\sigma_s}e_{i,\sigma_i}\\ \nonumber
    &\times \int_{-a}^{0} \dd z\,\chi^{(2)}_{\alpha\beta\gamma}(z)\,E_{p,\gamma}(\mathbf{q}_s+\mathbf{q}_i,z)\\
    &\times g_{\sigma_s\alpha}(\mathbf{q}_s,z_s,z,\omega_s)\,g_{\sigma_i\beta}(\mathbf{q}_i,z_i,z,\omega_p-\omega_s)\,.
\end{align}
Here, equation (\ref{eq:JSA_SI}) is understood as a joint angular probability (JAP) amplitude. Equation (\ref{eq:prob_SI_fourier2}) represents the detection rate for signal and idler photon-pairs generated through SPDC in a transversally homogeneous nonlinear source, accounting for spatial coordinates, frequencies, and detection polarization. Notably, $\mathcal{R}$ in the form of (\ref{eq:prob_SI_fourier2}) resembles a Fourier transform integral, enabling Fourier analysis for interpreting the probability in the more intuitive angular domain. 

However, since our focus is mainly on the far-field properties of the generated photon pairs, we will apply a far-field approximation to the probability rate similar to the classical Fraunhofer approximation for field diffraction \cite{born}. We aim to detect both photon pairs in the far field in medium 1 with refractive index $n_1$. Here, $k_1=\frac{\omega}{c}n_1$ is the wave number of plane waves in medium 1, and $k_{z,1}=\sqrt{k_1^2-k_x^2-k_y^2}$ denotes the $z$-component of its wave vector $\mathbf{k}_1=\mathbf{q}+k_{z,1}\hat{z}$. Note that $k_x$ and $k_y$ are the transversal components of the wave-vector for a photon generated in medium 2, which are also conserved in medium 1 due to the system's transversal invariant nature.

The far-field approximation requires joint detection of photons at sufficiently long distances from the source, i.e., $\mathbf{r}_s\gg\mathbf{r}$ and $ \mathbf{r}_i\gg\mathbf{r}$. Additionally, we assume a finite-sized pump beam, implying that detection occurs at a much greater distance compared to the extent of the pump illumination across the nonlinear slab (see appendix \ref{ap:FF-appx} for details).

Under this approximation the joint detection rate in Eq.~(\ref{eq:prob_SI_fourier2}) is simplified to (see appendix \ref{ap:FF-appx})
\begin{align}\label{eq:far_prob_IS}
    &\mathcal{R}^{(\text{far})}(\theta_s,\theta_i,\varphi_s,\varphi_i,\mathbf{e}_s, \mathbf{e}_i,\omega_s,\omega_p-\omega_s)=\frac{8}{\pi}\frac{1}{(2\pi)^4} \nonumber \\& n_in_s\frac{(\omega_p-\omega_s)^3\omega_s^3}{c^6} 
    \,\bigg|\Tilde{\mathcal{R}}(\mathbf{q}_s,\mathbf{q}_i,\omega_s,\omega_p-\omega_s)\;k_{z,1s}k_{z,1i}\bigg|^2\,.
\end{align}
Here, $\mathbf{q}_s=k_{x,s}\hat{x}+ k_{y,s}\hat{y}$, where $k_{x,s}=k_{1s}\sin\theta_{s}\cos\varphi_{s}$ and $k_{y,s}=k_{1s}\sin\theta_{s}\sin\varphi_{s}$, refer to the transverse k-vector of the signal photon detected in the far-field at propagation angles $\{\theta_s,\varphi_s\}$, and similarly for the idler photon. Thus, Eq. (\ref{eq:far_prob_IS}) describes the joint detection rate in the far-field as a function of the detection angles $\{ \theta_s$, $\theta_i$, $\varphi_s$, $\varphi_i\}$ in spherical coordinates.  Note that $k_{z,1s}$ and $k_{z,1i}$ in Eq. (\ref{eq:far_prob_IS}) correspond to the detection medium 1. If the detection medium changes, these factors should be adjusted accordingly (see appendix \ref{ap:FF-appx}). Also note that, $\tilde{\mathcal{R}}$ in Eq. (\ref{eq:far_prob_IS}) is $z$-independent, where
$
\Tilde{\mathcal{R}}(\mathbf{q}_s,\mathbf{q}_i,z_s,z_i, \omega_s,\omega_p-\omega_s)=\Tilde{\mathcal{R}}(\mathbf{q}_s,\mathbf{q}_i, \omega_s,\omega_p-\omega_s)e^{\ii k_{z,1s}z_s}e^{\ii k_{z,1i}z_i}    \,
$ (see appendix \ref{ap:FF-appx}).

In the far-field, the joint detection rate is proportional to the squared absolute value of $\Tilde{\mathcal{R}}$ multiplied by the signal and idler $z$-components of the wave vector, $k_{z,1s}$ and $k_{z,1i}$. In the paraxial regime ($q\ll k_1$ with $q=\abs{\mathbf{q}}$), $k_{z,1s}$ and $k_{z,1i}$ tend to $k_{1s}$ and $k_{1i}$, resulting in the far-field joint detection rate being directly proportional to the squared absolute value of $\Tilde{\mathcal{R}}$. This calculation extends the concept of classical Fraunhofer approximation \cite{born} to two-photon joint-detection. As our model is capable of handling emission angles up to 90°, it highlights the importance of the factors $k_{z,1s}$ and $k_{z,1i}$ in the nonparaxial regime.
The importance of the longitudinal wave-vectors in nonparaxial description of quantum-light propagation were also shown in studying the fundamental resolution limit of quantum imaging schemes \cite{vega2022}. 
Equations (\ref{eq:prob_SI_fourier2}), (\ref{eq:JSA_SI}), and (\ref{eq:far_prob_IS}) constitute the main results of this section, to be numerically solved in section (\ref{sec:farfield}).

\section{Green's Function of a Dielectric Slab}
\label{sec:GFs}

The optical properties of the system at the signal and idler wavelengths are fully contained in their respective Green's functions (GFs). To implement Eq. (\ref{eq:far_prob_IS}), we employ a GF suitable for the multilayered scheme depicted in Fig. (\ref{fig:slab}). In this scenario, the GFs for signal and idler, $\tens{\mathbf{G}}(\mathbf{r},\mathbf{r}_s)$ and $\tens{\mathbf{G}}(\mathbf{r},\mathbf{r}_i)$ connect a point $\mathbf{r}$ within the $\chi^{(2)}$-nonlinearity region (medium 2, see Eq. (\ref{eq:prob_SI})) to the detection positions $\mathbf{r}_s$ and $\mathbf{r}_i$ in medium 1.

If a plane wave of the form $\mathbf{E}(\mathbf{r})=\mathbf{E}e^{\ii\mathbf{k}\cdot\mathbf{r}}$, with $\sqrt{\mathbf{k}\cdot\mathbf{k}}=k=\frac{\omega}{c}n$ satisfies Maxwell's equation in a homogeneous medium of refractive index $n$, two possible wave vectors $\mathbf{k}$ for a given $\mathbf{q}$ are obtained, representing upward and downward propagation. These wave vectors take the form $\mathbf{k}_+=\mathbf{q}+k_z\hat{z}$ (upward) and $\mathbf{k}_-=\mathbf{q}-k_z\hat{z}$ (downward). When $n$ is real and $q>n\omega/c$, $k_z$ becomes imaginary, resulting in an evanescent wave. Additionally, at a given $\mathbf{q}$, a solution can be either $s$-polarized (transverse electric) or $p$-polarized (transverse magnetic), defined by unit polarization vectors $\hat{s}$ and $\hat{p}$ as
 \begin{align}\label{eq:s}
    \hat{s}=&\frac{\hat{z}\cross\mathbf{k}_\pm}{\abs{\hat{z}\cross\mathbf{k}_\pm}}=\frac{1}{q}(-k_y\hat{x}+k_x\hat{y})\;,\\ \label{eq:p}
    \hat{p}_\pm=&\frac{\mathbf{k}_\pm}{\abs{\mathbf{k}_\pm}}\cross\hat{s}=\frac{1}{k}\bigg(\mp\frac{k_zk_x}{q}\hat{x}\mp\frac{k_zk_y}{q}\hat{y}+q\hat{z}\bigg)\,.
\end{align}
These vectors are normalized according to $\hat{s}\cdot\hat{s}= 1$ and $\hat{p}_\pm\cdot
\hat{p}_\pm= 1$. Note that $\hat{s}$ is real, rests on the $x$-$y$ plane, and is medium
independent, while $\hat{p}$ can be complex and is medium dependent. Also, $\hat{s}$ does not depend on the direction of $\mathbf{k}$, while $\hat{p}$ changes for upward- or downward-generated fields. 

Using the $s$- $p$-polarized fields, the GF in Fourier domain for such a system, where the field is generated in medium 2 and propagates outside the slab to medium 1, can be constructed as \cite{Sipe,lau2007}
\begin{align}\label{eq:gf+}
    \nonumber&\tens{g}_{21}(\mathbf{q},z,z',\omega)=\frac{-\ii}{2k_{z,2}}\bigg(T_{21}^{(s)}(q,z',\omega)\;\hat{s}\hat{s}+\\&
    T_{21}^{(p+)}(q,z',\omega)\;\hat{p}_{1+}\hat{p}_{2+}+T_{21}^{(p-)}(q,z',\omega)\;\hat{p}_{1+}\hat{p}_{2-}\bigg)e^{\ii k_{z,1}z}\,,
\end{align}
where $T_{21}(q,z',\omega)$ is the generalized transmission coefficient of
the $s$- and $p$-polarized fields generated from the source plane at $z'\in\{-a,0\}$ and transmitted to medium 1 (see appendix \ref{ap:GFs} for analytical expression). The unit vector $\hat{s}$ is parallel to the interface and $\hat{p}$ is perpendicular to the wave vector $\mathbf{k}$ and $\hat{s}$. The subscript numbers in $k_z$ indicate the corresponding medium.


The GF in Eq. (\ref{eq:gf+}) describes from right to left (apart from the phase factor $e^{\ii k_{z,1}z}$) the contributions of a downward $p$-polarized wave generated in medium 2 that transform into an upward $p$-polarized wave in medium 1, an upward $p$-polarized wave in medium 2 that transform into an upward $p$-polarized wave in medium 1 and both the contribution of the upward/downward $s$-polarized waves generated in medium 2 that transfers to medium 1. All of them after multiple reflections and transmission, whose information is contained in the coefficients $T_{21}$. 

\section{Pump Field with Multiple Reflections}
\label{sec:pump}

The pump field undergoes diffraction and multiple reflections within the slab. Illustrated in Fig. (\ref{fig:slab}), the pump originating from medium 3 propagates to $z=-a$, enters the slab, and reflects multiple times. Consequently, the amplitude of the pump field within medium 2 varies with $z$ and is influenced by the slab's thickness $a$.

In many theoretical calculations, it is common to assume a pump beam polarization parallel to the interface (e.g. $x$-polarized) for SPDC modeling. However, true transverse electromagnetic waves are idealizations and physical beams of light also contain a longitudinal polarization component. This component may influence the photon-pair generation process if the material properties permit such excitation. For our calculations, we consider a pump field polarized in the $x$-direction right before the interface ($z=-a$) with a longitudinal component in the $z$-direction, given by
\begin{align}
 \mathbf{E}_{p}(\mathbf{q},&z=-a)= \nonumber \\
 &E_{p,x}(\mathbf{q},z=-a)\hat{x}-E_{p,x}(\mathbf{q},z=-a)\frac{k_x}{k_{z,3p}}\hat{z}\,,   
\end{align}
where $k_{z,3p}$ represents the $z$-component of the pump wave vector in medium 3, and $E_{p,z}$ is calculated from $E_{p,x}$ using the transversality constraint $\mathbf{k}_p\cdot\mathbf{E}_p=0$ (see Appendix \ref{ap:pumpCalc}).

Taking $\mathbf{E}_{p}(\mathbf{q},z=-a)=U_p(\mathbf{q},z=-a)\mathbf{\hat{e}}(\mathbf{q})$ one finds $\mathbf{\hat{e}}(\mathbf{q})={k_{z,3p}}/{\sqrt{k_{3p}^2-k_y^2}}\;\hat{x}-{k_x}/{\sqrt{k_{3p}^2-k_y^2}}\;\hat{z}$, which allows us to express the pump field in the form
\begin{align}\label{eq:pump}
    \nonumber \mathbf{E}_{p}&(\mathbf{q},z=-a)=\\
    &U_p(\mathbf{q},z=-a)\left[\frac{k_{z,3p}}{\sqrt{k_{3p}^2-k_y^2}}\hat{x}-\frac{k_x}{\sqrt{k_{3p}^2-k_y^2}}\hat{z}\right]\,,
\end{align}
where we take $\mathbf{\hat{e}}=\hat{x}$ for $k_x=k_z=0$. $U_p(\mathbf{q},z=-a)$ is the angular spectrum of the pump field at the interface. We assume it follows a Gaussian distribution,  
\begin{equation} \label{eq:pump_spec}
    U_{p}(\mathbf{q},z=-a)=
    \begin{cases}
       Ae^{-\mathbf{q}^2/w^2}\,,&\quad q^2\leq k_{3p}^2\\
       0\,, &\quad\text{otherwise,}
     \end{cases}  
\end{equation}
where $w$ is the width of the Gaussian angular spectrum. The condition $q^2\leq k_{3p}^2$ ensures that Eq.~(\ref{eq:E_p_fourier}) comprises only propagating waves in medium 3. Moreover, the spectrum width influences the contribution of the $z$-component of the pump beam: larger $w$ results in greater $E_{p,z}$. This is especially relevant in tightly focused pump beam scenarios (refer to Appendix \ref{ap:pumpGaus} for details).

At the interface, the pump beam can also be described using $s$- and $p$-polarized fields. After transmission into the slab, the $z$-dependent pump beam can be calculated as (see Appendix \ref{ap:pumpCalc} for details)
\begin{align}\label{eq:Epump}
    \nonumber &\mathbf{E}_p(\mathbf{q},z)=Q^{(s)}(q,z)E^{(s)}_p(\mathbf{q},z=-a)\hat{s}\\
    &+E^{(p)}_p(\mathbf{q},z=-a)\left(Q^{(p+)}(q,z)\hat{p}_{2+}+Q^{(p-)}(q,z)\hat{p}_{2-}\right)\,,
\end{align}
where $Q^{(s)}(q,z)$ is the transmission coefficient from medium 3 to medium 2, while $E^{(s)}_p(\mathbf{q},z=-a)$ and $E^{(p)}_p(\mathbf{q},z=-a)$ are $s$- and $p$-components of the pump before the slab. Unit vectors $\hat{s}$, $\hat{p}_{2+}$, and $\hat{p}_{2-}$ are defined in Eqs. (\ref{eq:s}, \ref{eq:p}) for medium 2.

\section{Far-Field Joint-Radiation Properties}
\label{sec:farfield}

\subsection{Degenerate photon-pair detection}

Now that we have the model, let us consider a Zinc-Blende GaAs crystal (medium 2) with thickness $a$ and orientation $\left\langle100\right\rangle$. The crystal axes are aligned with the laboratory coordinates $\{x, y, z\}$. The GaAs is on a SiO$_2$ substrate (medium 3, assumed semi-infinite), with air as medium 1 for detection. The pump from medium 3 is described by Eq. (\ref{eq:pump}) with $\lambda_p=500$ nm and Gaussian angular spectrum (Eq. (\ref{eq:pump_spec})) with $w=6.6\times10^5$ m$^{-1}$ (corresponding to a beam waist of $W=3\,\mu$m). At degenerate wavelengths ($\lambda_s=\lambda_i=1 \,\mu$m), we have $\epsilon_{2s}=\epsilon_{2i}=12.06$ in GaAs, while the pump relative permittivity at $\lambda_p=500$ nm in GaAs is $\epsilon_{2p}=17.63+3.83\ii$ \cite{papatry2021}. In SiO$_2$ substrate, $\epsilon_{3p}=2.14$ and $\epsilon_{2s}=\epsilon_{2i}=2.10$ \cite{Malitson65}, while in the detection medium, $\epsilon_{1p}=\epsilon_{1s}=\epsilon_{1i}=1$. The non-vanishing elements of the second-order susceptibility tensor of GaAs are $\chi^{(2)}_{yzx}=\chi^{(2)}_{zyx}=\chi^{(2)}_{xzy}=\chi^{(2)}_{zxy}=\chi^{(2)}_{xyz}=\chi^{(2)}_{yxz}=\chi^{(2)}_0$.

\begin{figure*}[ht]
\centering
\includegraphics[scale=1]{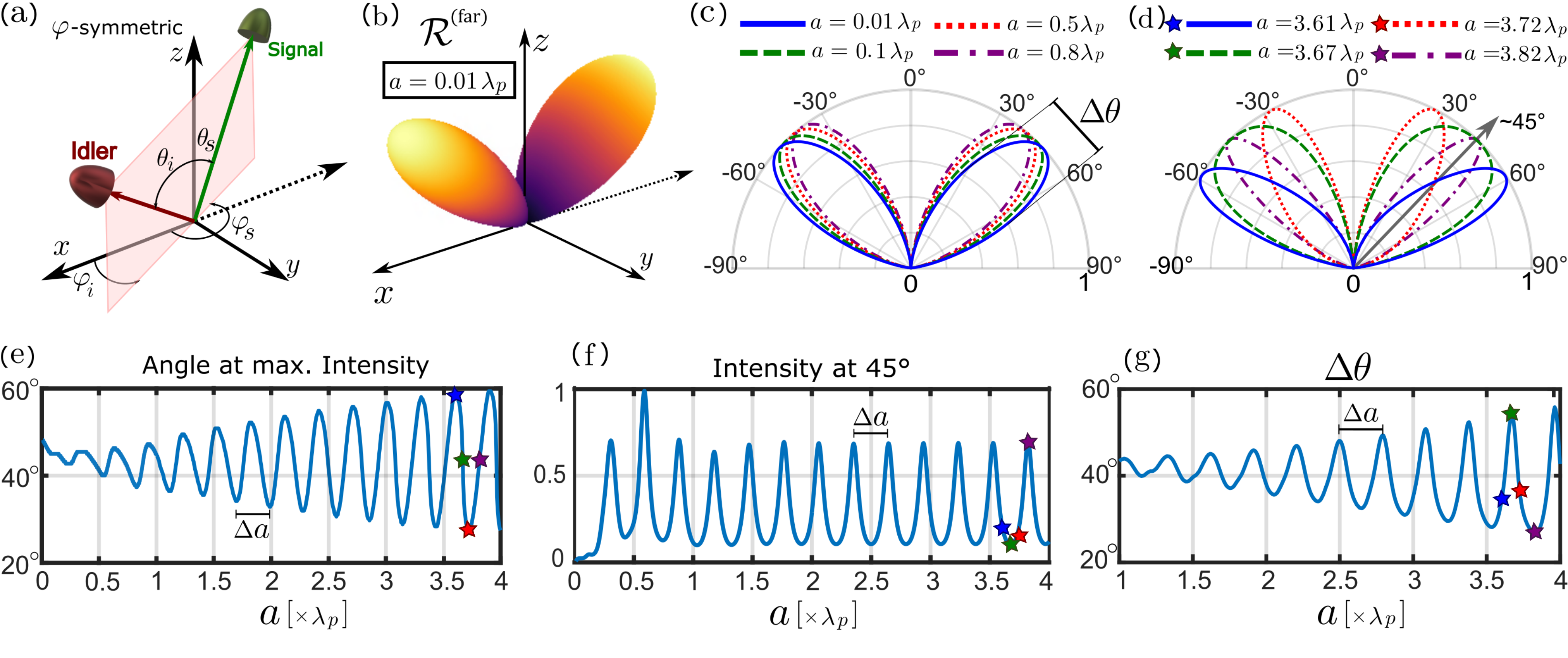}
\caption{ (a) $\varphi$-symmetric configuration for joint detection, which is considered for all the results shown in this figure. (b)  far-field joint detection probability for an ultra-thin film of GaAs of thickness $a=0.01\lambda_p$ in the $\varphi$-symmetric configuration. (c,d) slice cut through far-field radiation pattern in the $xz$ plane for different thicknesses (below and above $a=\lambda_p$) of the slab with normalized intensities. (e) angle of maximum joint emission as a function of the thickness of the slab. The four values plotted in (d) are marked by the stars, corresponding to the peak (blue star at $\sim 58.5$°), two in the middle (green and purple stars at $\sim 45$°) and the valley (red star at $\sim 27.5$°). (f) Intensity of joint detection at $\theta_s=45$° which exhibit FP oscillations. (g) Angle uncertainty $\Delta\theta$ of joint emission as a function of the thickness of the slab. The values in (d) are also in (f) and (g). The period of oscillation $\Delta a\sim0.29\lambda_p$.}
\label{fig:2}
\end{figure*}

We start by computing the coincidence detection rate $\mathcal{R}^{(\text{far})}$ in Eq.~(\ref{eq:far_prob_IS}) for unpolarized joint-detection. This involves summing over the three polarization directions of each signal and idler detector ($\mathbf{e}=\hat{x}$, $\hat{y}$, $\hat{z}$), resulting in nine combinations. Due to the relatively narrow angular range of our pump beam, opposite transverse wave vectors ($\mathbf{q}_s\sim-\mathbf{q}_i$) significantly contribute to the generation, as consequence of the transverse phase-matching condition ($\mathbf{q}_p=\mathbf{q}_s+\mathbf{q}_i$). For degenerate photon-pairs, this motivates investigating $\mathcal{R}^{(\text{far})}$ in terms of propagation angles $\{\theta_s,\varphi_s\}$ for the signal photon, where the idler is detected symmetrically at $\{\theta_i=\theta_s,\varphi_i= \pi +\varphi_s\}$. This configuration is shown in Fig.~\ref{fig:2}(a) and is called $\varphi$-symmetric throughout the manuscript. The $\varphi$-symmetric setup has already been proven crucial for entangled state generation in nanoresonators \cite{weissflog2024} and will also show to be essential for thin films in this work. 


The coincidence rate  in the $\varphi$-symmetric configuration is shown in Fig. \ref{fig:2}(b) for an ultra-thin source of thickness $a=0.01\lambda_p$, emphasizing again that both photons are detected in medium 1. Here, the $\chi^{(2)}_{yzx}$ and $\chi^{(2)}_{zyx}$ dominate the pair generation process due to weak $y$- and $z$-polarized and dominant $x$-polarized components of the pump field inside the slab. Note that $y$-components of the pump field are created inside the slab due to refraction, despite being zero outside (See Eq. (\ref{eq:Epump})). Hence, all components of the GaAs nonlinear tensor participate in the generation process. Yet, the dominance of $\chi^{(2)}_{yzx}$ and $\chi^{(2)}_{zyx}$ components implies that signal and idler fields within the slab are mainly generated by $y$- and $z$-polarized point-like nonlinear sources, resulting in no radiation in the $y$-$z$ plane, as observed in Fig. \ref{fig:2}(b). Additionally, emission at large angles is significantly reduced, despite the GaAs nonlinear tensor favoring stronger emission near $\theta_s=90$° for such an ultra-thin crystal. This reduction is mainly due to the strong decrease in transmittance from inside to outside of the slab at larger emission angles, ultimately reaching zero for outside emission angles of $\theta_s=90$° due to total internal reflection experienced by photons inside the slab.

In Fig. \ref{fig:2}(c), we display cuts along the $xz$-plane for various subwavelength thicknesses of the slab, $a=0.01\lambda_p$, $a=0.1\lambda_p$,  $a=0.5\lambda_p$ and $a=0.8\lambda_p$. Fig. \ref{fig:2}(d) illustrates $xz$-plane cuts for larger slab thicknesses, $a=3.61\lambda_p$, $a=3.67\lambda_p$, $a=3.72\lambda_p$ and $a=3.82\lambda_p$, which are special points for understanding the Fabry-Pérot dynamics, marked in Figs. \ref{fig:2}(e-g). All graphs are normalized to one for better observation of angular dependencies. Fig. \ref{fig:2}(c) reveals minimal angle variations in joint emission radiation for different subwavelength thicknesses, primarily influenced by the nonlinear tensor structure and angle-dependent transmission at interfaces. In contrast, Fig. \ref{fig:2}(d) demonstrates more significant angular changes for larger slab thicknesses. This is mainly due to the Fabry-Pérot effect, which enhances or reduces radiation in some angles due to constructive or destructive interferences, respectively. Notably, the angular pattern is highly sensitive to slab thicknesses in this regime.
To better understand the dynamics caused by Fabry-Pérot effect, we show in Figs. \ref{fig:2}(e) the angle of maximum intensity, (f) the intensity at $\theta_s=45$°, and (g) and the angular width of the emission pattern $\Delta\theta$, respectively, all as a function of the slab thickness. 
In Fig. \ref{fig:2}(e), the maximum emission angle exhibits a growing oscillatory pattern with increasing the thickness. 
The oscillatory behavior occurs around a mean angle of $\sim 45$°. 
At the detection angle of $\theta_s=45$°, we plot in Fig. \ref{fig:2}(f), the detection intensity (normalized to maximum) as a function of the slab thickness. We clearly observe an etalon-like effect caused by the FP interferences, where peaks of intensity repeat periodically with increasing slab thicknesses, with a period of $\Delta a\sim0.29\lambda_p$. This period of oscillation is the same on all three plots of Figs. \ref{fig:2}(e-g).
The angular uncertainty of emitted photons exhibits a similar thickness-dependent variation. We define the angle uncertainty $\Delta\theta$ as the difference between the angles at 50$\%$ of the maximum intensity (full width at half maximum). In Fig. \ref{fig:2}(g), we again observe increasing oscillations for thicknesses above $a=\lambda_p$, while it remains relatively constant for subwavelength thicknesses, as also seen in Fig. \ref{fig:2}(c).

It can be verified, that the peaks and valleys in Fig. \ref{fig:2}(f), correspond to constructive and destructive interference conditions, satisfying the relations $\Delta=m\lambda_s$ and $\Delta=(m+\frac{1}{2})\lambda_s$, respectively, where $m$ is a positive integer. $\Delta=2\,a\,n_{2s}\cos(\theta_{2s})$ is the optical path difference (OPD) between successive plane waves that are transmitted to medium 1 (being air) after a round trip in the slab \cite{lipson}. $\theta_{2s}$ is the angle in medium 2 such that it yields $\theta_{1s}=45$° in medium 1 after refraction.
The period of the resonance peaks is found by $\Delta a=\frac{\lambda_s}{2 \,n_{2s}\cos(\theta_{2s})}$, which results in $\Delta a\approx 0.29\lambda_p$, which matches the numerical results.
These points of constructive and destructive interferences for $\theta_{1s}=45$° also coincide with peaks and valleys of $\Delta\theta$, see in Fig. \ref{fig:2}(g). 
We point out that the angles of constructive interference oscillate around the $\theta_{1s}=45$°, as the thickness is increased, which results in an oscillation of the angles of maximum intensity around $\theta_{1s}=45$° in Fig. \ref{fig:2}(e). This can also be seen in Fig. \ref{fig:2}(d), where at thicknesses of $a=3.61\lambda_p$ or $a=3.72\lambda_p$, the directionality plots are maximized at angles $\sim58.5$° or $\sim27.5$°, respectively. 

A crucial observation is that due to the inherent losses in GaAs at $\lambda_p=500$ nm, there are constrictions on longitudinal phase-matching effect. Calculating the pump's decay length inside the GaAs slab, denoted as $\alpha_p=1/k''_{2p}$, where $k_{2p}=k'_{2p}+\ii\,k''_{2p}$, yields $\alpha_p=0.35\lambda_p$. When compared with the coherence length of our system under normal incidence, $L_c=\pi/\Delta k_{\parallel}$, where $\Delta k_\parallel=k'_{2p}-k_{2s}\cos\theta_{2s}-k_{2i}\cos\theta_{2i}$, we obtain $L_c=0.6\lambda_p$ for joint detection in medium 1 at $\theta_{1s}=\theta_{1i}=45$°. Notably, the decay length is smaller than the coherence length ($\alpha_p<L_c$), which is important, as it limits the range of impact of longitudinal phase matching with increasing slab thicknesses. Consequently, the FP effect at the signal and idler wavelengths has a dominant effect in creating the oscillatory behaviour observed in Figs.~\ref{fig:2}(e-g), and not the longitudinal phase matching.
Hence, the longitudinal phase matching, only affects the height of the first few peaks in Fig.~\ref{fig:2}(f), while subsequent peak intensities stabilize.

\subsection{Non-degenerate photon-pair detection}

Now we consider a frequency non-degenerate scenario for photon-pair detection. Unlike the degenerate case, the $\varphi$-symmetric configuration is not expected to dominate in the non-degenerate case.
For a normally incident plane-wave pump, with $\mathbf{q}_p=0$, the emitted pairs satisfy the transversal phase-matching condition of $\mathbf{q}_s=-\mathbf{q}_i$. This right away sets $\varphi_i=\varphi_s+\pi$ for signal and idler photons in a pair. Then, taking $\theta$ as the emission angle with respect to the forward $z$-direction we get $k_{x}^2+k_{y}^2=k^2\sin^2\theta$ for both signal and idler. Using the degeneracy factor $r$ defined as $r\equiv\lambda_s/\lambda_i$, we get $|\sin\theta_{s}|=r\abs{n_i/n_s}|\sin\theta_{i}|$ as the relation between the angles of emission for signal and idler photons in a pair. In medium one, we have $n_i=n_s=1$. 

\begin{figure}[ht]
\centering
\includegraphics[scale=1]{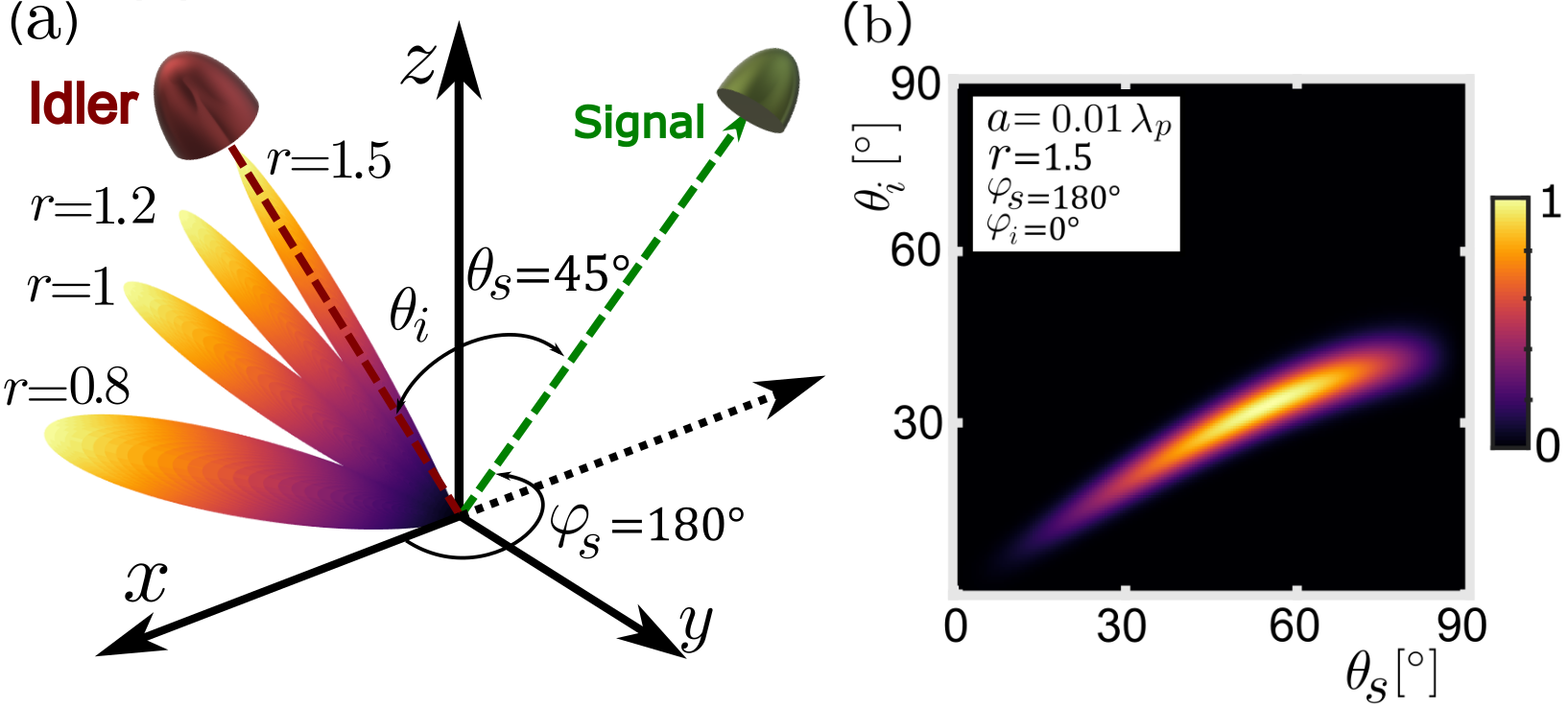}
\caption{(a) Coincidence rate for different degeneracy factors $r$ for a fixed signal detector position at $\theta_s=45$°,$\varphi_s=180$° and varying idler detector positions. The maximum angle of detection in the plot are $\theta_i(r=0.8)\sim61.2$°, $\theta_i(r=1)\sim45$°, $\theta_i(r=1.2)\sim36.3$° and $\theta_i(r=1.5)\sim28.2$° (b) Joint detection probability as a function of $\theta_s$ and $\theta_i$ in the $xz$-plane (with fixed $\varphi_s=0$ and $\varphi_i=180$°) for $r=1.5$. The thickness is $a=0.01\lambda_p$ in both graphs.}
\label{fig:3}
\end{figure}

In Fig. \ref{fig:3}(a), we calculate the joint detection probability with a fixed signal detector at $\theta_s=45$°, $\varphi_s=180$° for nondegeneracy factors $r=0.8$, $1$, $1.2$, and $1.5$ in the ultra-thin case ($a=0.01\lambda_p$). Here, smaller idler wavelengths ($r>1$) tend to emit the idler closer to the optical axis, while larger idler wavelengths ($r<1$) push the idler away from the optical axis. Maximum emission angles calculated using the relation between angles are $\theta_i(r=0.8)=62.11$°, $\theta_i(r=1)=45$°, $\theta_i(r=1.2)=36.1$° and $\theta_i(r=1.5)=28.12$°, matching Fig. \ref{fig:3}(a). For $r=1$ the maximum coincidence rate lie along $\theta_s=\theta_i$, justifying the previously used $\varphi$-symmetric configuration. Additionally, in Fig. \ref{fig:3}(a), we observe that as the degeneracy factor $r$ decreases, the uncertainty in idler angles increases, meaning weaker correlations in emitted photons. 

Due to transverse phase matching, most pairs are generated under the condition $\varphi_s=\varphi_i+180$°, which suggests visualizing the joint-detection rate as a function of the angles $\{\theta_s,\theta_i\}$, in a particular plane defined by $\varphi$. In Fig. \ref{fig:3}(b) we plot $\mathcal{R}^{(\text{far})}$ along the $xz$-plane, where $\varphi_i=0$ and $\varphi_s=180$°, as a function of $\theta_s$ and $\theta_i$ for $r=1.5$ and $a=0.01\lambda_p$. Here we can see that a better correlation in the non-degenerate case can be achieved by detecting at small angles, however this comes at expenses of the efficiency of the process. Additionally, emission at $\sim45$°, dominant in the degenerate case, is no longer prevalent in the non-degenerate scenario, which is now shifted by transverse phase-matching.
In the depicted scenario in Fig.~\ref{fig:3}(b), these angles are $\theta_s=55.35$° and $\theta_i=33.3$°.

\subsection{Absolute pair generation rates}

A critical question revolves around the overall efficiency of the pair-generation process. The total number of pairs that can be collected by a lens of numerical aperture NA, is obtained by integrating the differential pair-rate $\mathcal{R}\equiv\frac{\dd^4N}{\dd t\dd\Omega_s\dd\Omega_i\dd\omega_s} $ over the entire signal and idler solid angles that fall within the angle $\theta=\arcsin(\text{NA})$ where $\dd\Omega=\sin\theta\dd\theta\dd\varphi$, then summing over all possible detected polarization combinations, and finally integrating over the frequency range of the detected signal photons. To do this, we consider a $0.59\lambda_p\approx 295$ nm for the GaAs slab thickness, which corresponds to the maximum intensity peak in Fig. \ref{fig:2}(f). We assume a pump power of 1 mW spread over the Gaussian beam of width $W=3\mu$m with spectrum defined in Eq. (\ref{eq:pump_spec}), where the relation $W=2/w$ is satisfied in the paraxial regime. The incident pump power $P_{\mathrm{pump}}$ is related to the amplitude of the Gaussian spectrum through $A=\sqrt{\mu_0 \pi cP_{\mathrm{pump}} W^2}$, in the paraxial pump limit. For $\langle 100\rangle$ GaAs we take $\chi^{(2)} \approx 300$ pm/V as an approximate value for its nonlinearity coefficient based on different frequency-dependent measurements \cite{Shoji}.

After the solid-angle integrations, we obtain the quantity $\dd^2 N_{pair}/(\dd t\dd \omega_s)$, where we calculate this for the fixed signal frequency corresponding to the degenerate case, at which $\lambda_s=\lambda_i=1\,\mu$m. This quantity is dimensionless and represents the number of photon pairs per second that one can collect per units of signal photon frequency. For NA values of 0.6, 0.8 and 0.9 we obtain $\dd^2 N_{pair}/(\dd t\dd \omega_s)\approx 5.2\times10^{-13}$, $1.7\times10^{-12}$ and $2.6\times10^{-12}$, respectively. Assuming that the generation efficiency is more or less constant for about 10 nm of bandwidth for the signal photons (i.e. from 1 to 1.01 $\mu$m wavelength), which is a good approximation based on the result in Fig. \ref{fig:4}(b), we can multiply the derived $\dd^2 N_{pair}/(\dd t\dd \omega_s)$ by a corresponding $\Delta\omega\approx 1.87\times10^{13}$. Then, we find the total rate of photon pairs with a bandwidth of 10 nm for the signal photons, collected with NA values of 0.6, 0.8 and 0.9, to be about 10, 32, 49 Hz. These numbers are comparable to SPDC measurements in sub-micron-thick thin films \cite{Chek_2021,weissflog2023tunable}, once the measured rates are scaled appropriately for experimental collection efficiencies and detection bandwidths. Finally, we note that changing the width of the pump beam from $3\,\mu$m to $6 \,\mu$m (carrying the same 1mW of power), does not change the collected photon-pair rate appreciably, where with an NA$=0.9$ for the collection lens and 10 nm bandwidth for the signal photon, we find a photon-pair rate of 51 Hz.

\section{Quantum Polarization Tomography}
\label{sec:Qtomo}

So far we have examined the mechanism of photon-pair generation by a thin-film nonlinear source, concentrating on the far-field directionality properties as functions of the slab thickness and photon-pair wavelengths. For this purpose, our investigation has focused only on polarization-independent joint-detection. However, we are also interested in the polarization state of the photon pairs, which we will study in this section. Although our model based on the GF quantization approach does not directly predict the quantum state of the photon pairs, but rather predicts detection rates, it still allows us to make these detection rates dependent on the specific wavelength, position, and polarization of the detected photon pairs. Utilizing this, one can determine the density matrices of the biphoton polarization states following a tomographic method \cite{James,weissflog2024}. There, the density matrix $\hat{\rho}$, expressed in the $\{\ket{HH},\ket{HV},\ket{VH},\ket{VV}\}$ basis, is retrieved from projective measurements computed for a set of 16 tomographic states $\{\ket{\psi_\nu}\}_{\nu\in[[1;16]}$ \cite{James,weissflog2024}. The details of the tomographic approach can be found in the appendix \ref{ap:QPT}. 

\begin{figure}[ht]
\centering
\includegraphics[scale=1]{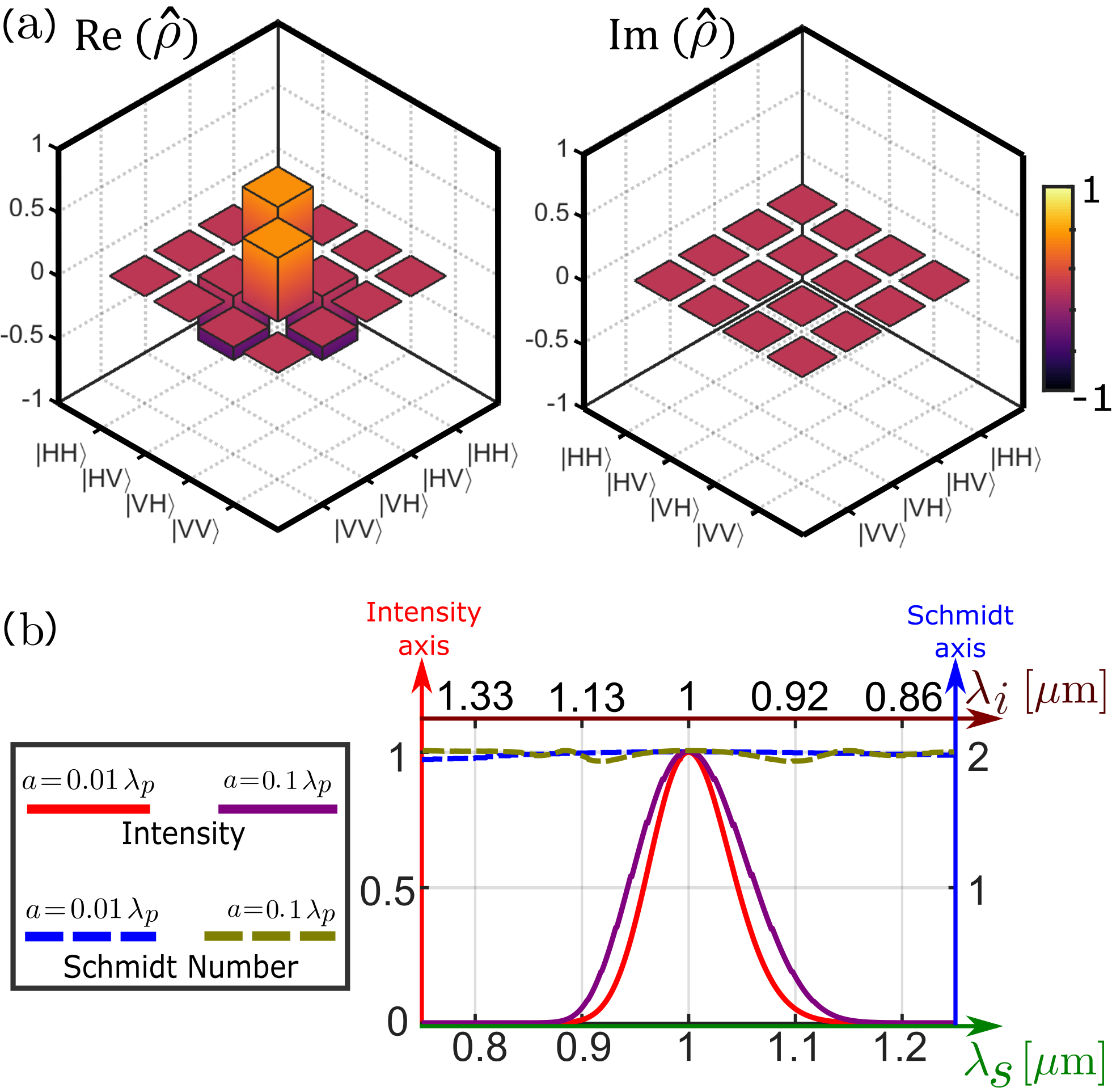}
\caption{(a) Real and imaginary parts of the elements of the density matrix, $\hat{\rho}$, for the polarization state of a degenerate ($r=1$) photon pair, generated from an ultra-thin GaAs slab of thickness $a=0.01\lambda_p$. The detectors are in a $\varphi$-symmetric configuration with $\theta_s=45$° and $\varphi_s=0$. (b) Schmidt number and joint-coincidence rate as a function of the signal wavelength (bottom axis) and idler wavelength (top axis) for the ultra-thin GaAs slab in the same $\varphi$-symmetric configuration, for two thicknesses $a=0.01\lambda_p$ and $a=0.1\lambda_p$}
\label{fig:4}
\end{figure}

Using this method, we examine the density matrix for several scenarios, starting with an ultra-thin slab ($a=0.01\lambda_p$) in a $\varphi$-symmetric configuration detection ($\theta_s=45$°, $\varphi_s=0$) for $r=1$. The pump is the same as in Sec.~(\ref{sec:farfield}). In Fig.~\ref{fig:4}(a), we display the real and imaginary parts of the density matrix $\hat{\rho}$ representing the polarization state. Remarkably, under these conditions, the density matrix corresponds to a maximally polarization-entangled state of the form $\ket{\psi}=\ket{H}_s\ket{V}_i-\ket{V}_s\ket{H}_i=\ket{HV}-\ket{VH}$. For every photon-pair state, we assess the level of polarization entanglement by determining the Schmidt number, $K$. A value of $K=1$ indicates a completely unentangled polarization state, while a value of $K=2$ means a fully entangled state (see appendix \ref{ap:QPT}). We investigate entanglement for two thicknesses ($a=0.01\lambda_p$ and $a=0.1\lambda_p$), including non-degeneracy, in Fig.~\ref{fig:4}(b) at the same $\varphi$-symmetric configuration. We plot both the Schmidt number ($K$) and the detection rate (normalized to one) as a function of the wavelength of the signal (bottom axis) and its corresponding idler photon (top axis). We observe that high-quality entanglement is preserved along an extremely broad range of wavelengths in both thicknesses, despite the joint-detection probability dropping at values $r\ne 1$ due to moving away from the transversal phase-matching condition (as in Fig. \ref{fig:3}(b)). This demonstrates GaAs's ability to produce polarization-entangled Bell states of the form $\ket{\psi}=\ket{HV}-\ket{VH}$ across a wide range of wavelengths, maintaining near-perfect entanglement at $\varphi$-symmetric detection with minor degradation at non-degenerate wavelengths.

\begin{figure}[ht]
\centering
\includegraphics[scale=1]{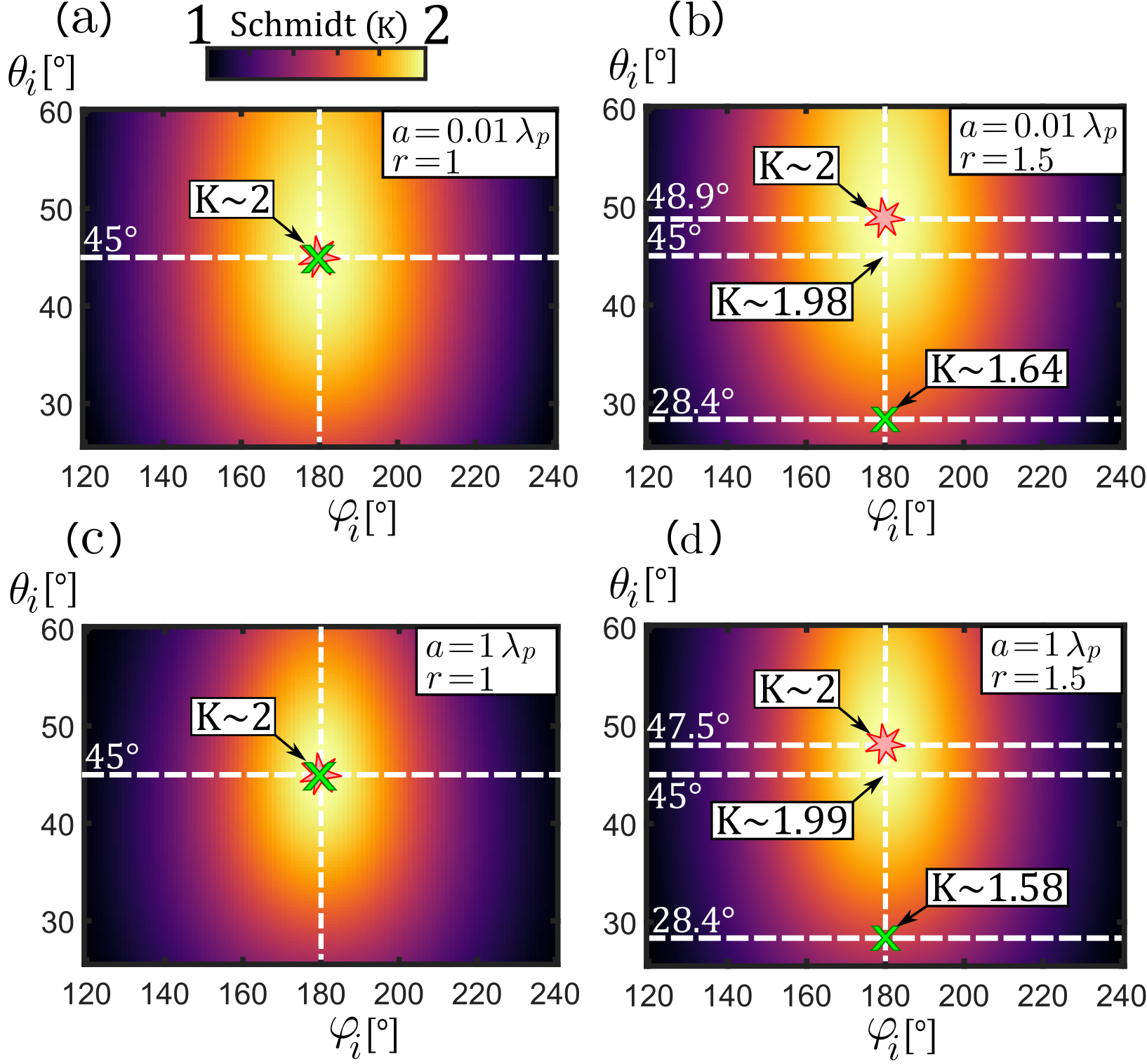}
\caption{Schmidt number as a function of the idler angles $\{\theta_i,\varphi_i\}$ with fixed $\theta_s=45$° and $\varphi_s=0$ for (a) $a=0.01\lambda_p$ and $r=1$, (b) $a=1\lambda_p$ and $r=1.5$, (c) $a=1\lambda_p$ and $r=1$, (c) $a=1\lambda_p$ and $r=1.5$. The angle of maximum entanglement and the angle of maximum detection rate are marked by a red "$*$" and a green "x", respectively.}
\label{fig:5}
\end{figure}

It is interesting to explore the preservation of broad entanglement obtained in the $\varphi$-symmetric setup across various emission angles, especially in non-degenerate scenarios where the $\varphi$-symmetric configuration is not dominant. Fig. \ref{fig:5} displays the Schmidt number ($K$) variation with $\{\theta_i,\varphi_i\}$, with fixed signal detection at $\theta_s=45$° and $\varphi_s=0$, for both degenerate (Fig. \ref{fig:5}(a, c)) and non-degenerate cases with $r=1.5$ (Fig. \ref{fig:5}(b, d)), with $a=0.01\lambda_p$ and $a=1\lambda_p$. Interestingly, while the angles of maximum entanglement (marked with red "$*$") and maximum joint-probability (marked with green "x") coincide for $r=1$, they move in opposite directions for $r=1.5$. Specifically, the angle of maximum detection rate approaches the optical axis rapidly, following transverse phase matching, whereas the angle of maximum entanglement shifts slightly away from the $\varphi$-symmetric angle. Notably, although entanglement slightly degrades in the non-degenerate $\varphi$-symmetric case (the values of $K\sim 1.98$ and $1.99$ in Figs. \ref{fig:5}(b, d)), there actually exists a close by angle at which the entanglement is maximum again at $K\sim 2$. Moreover, the angles of maximum detection rate and entanglement are significantly apart for the non-degenerate case,
indicating a trade-off between efficiency and entanglement with non-degenerate photon pairs, at least under the stated pumping conditions. Additionally, maximally polarization-entangled states exhibit strong robustness to changes in thickness and wavelengths, which parallels findings in Ref. \cite{weissflog2024}, where similar responses were observed in Bell state generation in GaAs nonlinear nanoresonators.

\subsection{Effects from a finite collection angle}

Throughout this work, we focused on an angularly-resolved analysis to best understand the emission dynamics of SPDC from thin films, which experimentally corresponds to detecting photons that are collected over infinitely narrow emission angles. In experimental scenarios, collection over extremely narrow angular ranges is not feasible, especially because it drastically reduces the observed number of photon pairs. Commonly, the two-photon emission from such nonlinear thin films is collected with an objective/lens that collects a finite range of emission angles \cite{Chek_2022,guo2023,weissflog2023tunable}. In the following, we investigate how collection over finite emission angles can affect the degree of entanglement.

We focus on a case with slab thickness of $a=0.01\lambda_p$, detecting only frequency-degenerate signal and idler photons. Our numerical analysis is performed in a way that resembles a simplified experimental scenario, in which a lens with numerical aperture NA is collecting all signal and idler photons going in any pairs of emission angles within the cone of angle $\theta=\arcsin(\mathrm{NA})$ (see appendix \ref{ap:QPT}), and the tomography is performed by integrating over the detection probabilities from these collected pairs. Firstly, we find that the polarization density matrices for the two-photon states are no longer fully pure as was before, but have a purity P$<1$, as shown in Table \ref{tab:NAs} for different values of collection NA and a fixed pump width of $W = 3\,\mu$m. This is due to the pump beam having a finite width, which makes the angular correlation between pairs not perfect (see Fig. \ref{fig:3}(a) for $r=1$), and hence in the process of integrating over the different collected angles, the information about these correlations is lost, which results in a mixed state. We verify this by finding the purity for a fixed NA and varying values of pump width, as shown in Table \ref{tab:Ws}. Here we see that larger pump widths, which produces higher angular correlations, result in a higher purity. Also, the fact that purity rises with larger NA in Tab. \ref{tab:NAs}, is attributed to the fact that, since in our system the pairs are mainly radiated around $\theta \sim 45$° (as shown in Fig. \ref{fig:2}), going towards higher NAs results in capturing the emission at angles near 45°, which have a dominant effect in the sum over all collected angles, thus improving the systems performance.

With a mixed state, a better measure for entanglement is the concurrence, C, a quantity ranging between C $= 0$ for separable and C $= 1$ for fully entangled states \cite{Wootters}. Here, we also observe a C$<1$, yet quite close to 1, which means that the entanglement degree is also slightly reduced. This can also be explained, as we showed in Fig. \ref{fig:5}, that the entanglement is only perfect for pairs of $\varphi$-symmetric angles, yet the fact that the pump has a finite width results in generation of degenerate pairs that deviate from this condition (Fig. \ref{fig:3}(a) for $r=1$). This means that we also collect pairs that are not perfectly polarization entangled, and this reduces the concurrence of the total collected state. This is verified from the data in Table \ref{tab:Ws}, which shows that the concurrence increases with a wider pump beam. Concurrence also improves with collection of larger angles, as seen in Table \ref{tab:NAs}, which we attribute to the same reason that purity increases with higher NA. Overall, we can see that with reasonable values of NA and pump widths, high values for purity and concurrence can be achieved, which can be improved consistently with increasing the pump width, at least for this given system. It is important to mention that the resultant entangled state is always close to $|\psi\rangle = |HV\rangle - |VH\rangle$ state.

\begin{table}[ht]
\centering
\caption{Purity and concurrence of the two-photon state for several collection numerical apertures (NAs), with a fixed pump width $W=3\,\mu$m. Signal and idler are frequency degenerate and slab thickness is $a=0.01\lambda_p$.}
\begin{tabular}{|l|l|l|l|}
Fixed $W=3\,\mu\mathrm{m}$ & $\mathrm{NA}=0.4$ & $\mathrm{NA}=0.6$ & $\mathrm{NA}=0.8$ \\ \hline
Purity          & $0.902$  & $0.954$  & $0.973$  \\ \hline
Concurrence     & $0.931$  & $0.968$  & $0.981$  \\ 
\end{tabular}
\label{tab:NAs}
\end{table}

\begin{table}[ht]
\centering
\caption{Purity and concurrence of the two-photon state for several pump widths $W$, with a fixed collection numerical aperture NA=0.4. Signal and idler are frequency degenerate and slab thickness is $a=0.01\lambda_p$.}
\begin{tabular}{|l|l|l|l|}
Fixed $\mathrm{NA}=0.4$ & $W=3\,\mu\mathrm{m}$ & $W=6\,\mu\mathrm{m}$ & $W=9\,\mu\mathrm{m}$ \\ \hline
Purity          & $0.902$  & $0.974$  & $0.989$  \\ \hline
Concurrence     & $0.931$  & $0.983$  & $0.992$  \\ 
\end{tabular}
\label{tab:Ws}
\end{table}

\section{Summary and Conclusion}
\label{sec:conc}

In summary, we have developed a theoretical framework based on the classical Green's function of the system, to comprehensively describe spontaneous parametric down-conversion (SPDC) in nonlinear thin films.
This vectorial formalism is capable of treating subwavelength slab thicknesses, accounts for Fabry-Pérot effects, and can address absorption in the slab. Additionally, we have introduced a simplified formulation for numerical investigation of far-field properties of photon pairs detected within one of the media surrounding the slab. Using a Zinc-Blende $\left\langle100\right\rangle$ GaAs crystal as a specific example, deposited on a SiO$_2$ substrate with air as the detecting medium, we have computed probability rates for both degenerate and non-degenerate emissions. Through analytical calculations and numerical simulations, we examined and explained the impact of Fabry-Pérot interferences, characterized by oscillations in emitted angles and probability distributions.

Furthermore, we explored the potential of this system to produce polarization-entangled states. Employing a tomographic approach, we demonstrated that a GaAs slab can generate maximally polarization-entangled states of the form $\ket{\psi}=\ket{HV}-\ket{VH}$, within a $\varphi$-symmetric configuration by analyzing its density matrix $\hat{\rho}$. Our findings revealed that the entanglement achieved remains highly resilient against variations in wavelengths and thickness, even in non-degenerate scenarios. However, in the case of non-degenerate photons, achieving maximal entanglement entails a trade-off with the efficiency of the process, where the latter is affected by the transversal phase-matching condition. This suggests a nuanced interplay between entanglement quality and detection rates in such systems, which could potentially be manipulated by using spatially structured pump fields to reach maximum entanglement together with high detection rates.

\section{acknowledgements}
The authors acknowledge funding by the German Ministry of Education and Research (13N14877, 13N16108), the Deutsche Forschungsgemeinschaft (DFG, German Research Foundation) through the Collaborative Research Center NOA (CRC 1375, project number 398816777), the German Federal Ministry of Education and Research BMBF through project QOMPLEX (13N15985), the Thuringian State Government through project Quantum Hub (project number 2021 FGI 0043), the Nexus program of the Carl-Zeiss-Stiftung (project MetaNN) and by the Deutsche Forschungsgemeinschaft (DFG, German Research Foundation) – Project number 505897284.

\appendix
\section{Far-Field Approximation of the Joint-Detection Probability }
\label{ap:FF-appx}
Let us consider a 2-dimensional inverse Fourier expansion integral of the form
\begin{align}\label{eq:gen}
    I(x,y,z)&=\nonumber\\&\frac{1}{(2\pi)^2}\iint \dd k_x\,\dd k_y\; U(k_x,k_y)\,\frac{e^{\ii k_z z}}{k_z}\,e^{\ii(k_xx+k_yy)}\,, \\
    \nonumber&=\mathcal{F}^{-1}\left [U(k_x,k_y)\,\frac{e^{\ii k_z z}}{k_z}\right]\,,
\end{align}
that describes the amplitude propagation of a field with angular spectrum $U(k_x,k_y)/k_z$ at the initial position $\mathbf{r}'=(x',y',z'=0)$ to the final position $\mathbf{r}=(x,y,z)$. With wave vector component in the $z$ direction $k_z=\sqrt{k^2-k_x^2-k_y^2}$.

To do a far-field approximation of this propagation integral, we first use the convolution theorem to write Eq. (\ref{eq:gen}) as a convolution of the form
\begin{align}
    I(x,y,z)&=\mathcal{F}^{-1}\left [U(k_x,k_y)\right]*\mathcal{F}^{-1}\left[\frac{e^{\ii k_z z}}{k_z}\right]\,,\\
    &= u(x,y)*\frac{-\ii}{2\pi}\frac{e^{\ii kr}}{r}\,,\\
    &=\frac{-\ii}{2\pi}\iint \dd x'\,\dd y'\,u(x',y')\,\frac{e^{\ii kR}}{R}\,, \label{eq:conv}
\end{align}
where $*$ denotes the convolution operation, $U(k_x,k_y)=\mathcal{F}[u(x,y)]$ and we have used the Weyl expansion of an outgoing spherical wave as a superposition
of plane waves, $ \frac{e^{\ii kr}}{r}=\frac{\ii}{2\pi}\iint\dd k_x\, \dd k_y\;\frac{e^{\ii k_z z}}{k_z}\,e^{\ii(k_xx+k_yy)} $. Where $r=\sqrt{x^2+y^2+z^2}$ is the magnitude of the position vector $\mathbf{r}$ and $R=\sqrt{(x-x')^2+(y-y')^2+z^2}$ is the magnitude of the vector $\mathbf{R}=\mathbf{r}-\mathbf{r}'$.

Eq. (\ref{eq:conv}) is interpreted as the beam propagation of an initial field $u(x',y')$ to a distance $z$ with an outgoing spherical wave as the response function $h={e^{\ii kr}}/{r}$. In the spherical coordinates representation with $x=r\sin\theta\cos\varphi$, $y=r\sin\theta\sin\varphi$ and  $z=r\cos\theta$, $R$ takes the form
\begin{align}
R=&r\bigg (1-2\sin\theta\cos\varphi\frac{x'}{r}-2\sin\theta\sin\varphi\frac{y'}{r}+\\ \nonumber 
&\frac{x'^2}{r^2}+\frac{y'^2}{r^2}\bigg)^{1/2}\,.
\end{align}
By taking the condition that $r\gg r'$ the last to terms are approximated to be zero, and using a first order binomial expansion of $R$, we arrive at
\begin{align}
R\approx  r\left(1-\sin\theta\cos\varphi\frac{x'}{r}-\sin\theta\sin\varphi\frac{y'}{r}\right)\,.
\end{align}
Now, when introducing $R$ into (\ref{eq:conv}), we notice that in the response function the phase factor varies rapidly in comparison to its denominator. Therefore we keep the 1st order expansion for the phase factor while using a 0th order as a good enough approximation for the denominator. Thus, we get the far-field-approximated expression for beam propagation,
\begin{align}
    I(r,\theta,\varphi)&=\frac{-\ii}{2\pi}\frac{e^{\ii kr}}{r}\iint \dd x'\,\dd y'\,u(x',y')\,\\ \nonumber
    &\times e^{-\ii(k\sin\theta\cos\varphi\,x'+k\sin\theta\sin\varphi\,y')}\,\\ \label{eq:far}
    &= \frac{-\ii}{2\pi}\frac{e^{\ii kr}}{r}\,U(k\sin\theta\cos\varphi,k\sin\theta\sin\varphi)\,.
\end{align}
This equation tells us that, under the far-field approximation, the amplitude propagation to the far-field of an initial beam field defined at plane $z=0$, with angular spectrum $U(k_x,k_y)/k_z$ is proportional to $U(k_x,k_y)$ with $k_x= k\sin\theta\cos\varphi$ and $k_y= k\sin\theta\sin\varphi$. 

This approximation, which is similar to the Fraunhofer approximation for the diffraction of classical beams \cite{born}, also considers nonparaxial propagation, meaning that it also describes propagation at angles up to 90°. However, our goal here is to extend this concept to the two-photon joint-detection probability. To this aim, we start with Eq.~(\ref{eq:prob_SI_fourier2}), and notice that the integral inside the absolute value can be written as
\begin{align}\label{eq:genR}
    \nonumber I_\mathcal{R}=&\frac{1}{(2\pi)^2}\int_{-\infty}^\infty\dd \mathbf{q}_s\;e^{\ii k_{z,1s}z_s}e^{\ii\mathbf{q}_s\cdot\mathbf{r}_{s\perp}}\\ \nonumber 
    &\times\frac{1}{(2\pi)^2}\int_{-\infty}^\infty \,\dd  \mathbf{q}_i\;\Tilde{\mathcal{R}}(\mathbf{q}_s,\mathbf{q}_i, \omega_s,\omega_p-\omega_s)e^{\ii k_{z,1i}z_i}\;\\
    &\times e^{\ii\mathbf{q}_i\cdot\mathbf{r}_{i\perp}}\,,
\end{align}
where $k_{z,1}$ is the $z$ component of the wave vector in the medium of detection (medium 1 in this case). The factors $e^{\ii k_{z,1s}z_s}$ and $e^{\ii k_{z,1i}z_i}$ come from the definition of the GF in Eq.~(\ref{eq:gf+}), which allows us to write 
\begin{align}
\Tilde{\mathcal{R}}(\mathbf{q}_s,\mathbf{q}_i,&z_s,z_i, \omega_s,\omega_p-\omega_s)=\nonumber \\ 
&\Tilde{\mathcal{R}}(\mathbf{q}_s,\mathbf{q}_i, \omega_s,\omega_p-\omega_s)e^{\ii k_{z,1s}z_s}e^{\ii k_{z,1i}z_i}    \,.
\end{align}

The integral in Eq. (\ref{eq:genR}) is a two-photon analogue of the propagation integral in Eq. (\ref{eq:gen}) and can be interpreted as the propagation of the probability amplitude of the generated signal and idler photon pair with joint angular probability (JAP) amplitude $\Tilde{\mathcal{R}}(\mathbf{q}_s,\mathbf{q}_i, \omega_s,\omega_p-\omega_s)$.

Therefore, we can perform a two-photon-like far-field approximation where we identify $\Tilde{\mathcal{R}}$ in Eq. (\ref{eq:genR}) as $U/k_z$ in Eq.~(\ref{eq:gen}), and by using the identity in (\ref{eq:far}) twice (for signal and idler propagation), we obtain
\begin{equation}\label{eq:farR}
    I_\mathcal{R}=\frac{-1}{(2\pi)^2} \frac{e^{\ii k_sr_s}}{r_s}\frac{e^{\ii k_ir_i}}{r_i} \Tilde{\mathcal{R}}(\mathbf{q}_s,\mathbf{q}_i, \omega_s,\omega_p-\omega_s)\,k_{z,1s}k_{z,1i}\,.
\end{equation}
Here, $\mathbf{q}_s=k_{x,s}\hat{x}+ k_{y,s}\hat{y}$, where $k_{x,s}=k_{1s}\sin\theta_{s}\cos\varphi_{s}$ and $k_{y,s}=k_{1s}\sin\theta_{s}\sin\varphi_{s}$, refer to the transverse k-vector of the signal photon detected in the far-field at propagation angles $\{\theta_s,\varphi_s\}$, and similarly for the idler photon. After inserting (\ref{eq:farR}) in (\ref{eq:prob_SI_fourier2}) we arrive at the expression in (\ref{eq:far_prob_IS}) which describes a far-field-approximated joint-probability rate of detecting coincidences in the far-field as a function of the detection angles $\{\theta_s,\theta_i,\varphi_s,\varphi_i\}$.

\section{Green's Function of a Dielectric Slab}
\label{ap:GFs}

The GFs in Fourier domain for the multilayered system depicted in Fig. (\ref{fig:slab}), where a field is generated in medium 2 and propagates outside the slab can be constructed as 
\begin{align}
    \nonumber&\tens{g}_{21}(\mathbf{q},z,z',\omega)=\frac{-\ii}{2k_{z2}}\bigg(T_{21}^{(s)}(q,z',\omega)\;\hat{s}\hat{s}+\\&
    T_{21}^{(p+)}(q,z',\omega)\;\hat{p}_{1+}\hat{p}_{2+}+T_{21}^{(p-)}(q,z',\omega)\;\hat{p}_{1+}\hat{p}_{2-}\bigg)e^{\ii k_{z1}z}\,,
\end{align}
and
\begin{align}
    \label{eq:gf-}
    \nonumber &\tens{g}_{23}(\mathbf{q},z,z',\omega)=\frac{-\ii}{2k_{z2}}\big[T_{23}^{(s)}(q,z',\omega)\;\hat{s}\hat{s}+\\&
    T_{23}^{(p-)}(q,z',\omega)\;\hat{p}_{3-}\hat{p}_{2-}+T_{23}^{(p+)}(q,z',\omega)\;\hat{p}_{3-}\hat{p}_{2+}\big]e^{-\ii k_{z3}z}\,,
\end{align}
where we have also added the GF connecting medium 2 and 3 for completeness.

Using the definitions for the $s$- and $p$-polarized fields in Eqs. (\ref{eq:s}) and (\ref{eq:p}) we can compute the dyadics $\hat{s}\hat{s}$, $\hat{p}_{1+}\hat{p}_{2\pm}$ and $\hat{p}_{3-}\hat{p}_{2\pm}$ as
\begin{align}
    \hat{s}\hat{s}&=\frac{1}{q^2} \left( \begin{matrix} k_y^2 & -k_xk_y & 0 \\ -k_xk_y & k_x^2 & 0 \\ 0 & 0 & 0 \end{matrix} \right)\;,\\
    \hat{p}_{1+}\hat{p}_{2\pm}&=\nonumber \\
    &\frac{1}{k_1k_2}\left( \begin{matrix} \pm\frac{k_x^2k_{z1}k_{z2}}{q^2} & \pm\frac{k_xk_yk_{z1}k_{z2}}{q^2} & -k_xk_{z1} \\ \pm\frac{k_xk_yk_{z1}k_{z2}}{q^2} & \pm\frac{k_y^2k_{z1}k_{z2}}{q^2} & -k_yk_{z1} \\ \mp k_xk_{z2} & \mp k_yk_{z2} & q^2 \end{matrix} \right)\,, \\
    \hat{p}_{3-}\hat{p}_{2\pm}&=\nonumber \\
    &\frac{1}{k_2k_3}\left( \begin{matrix} \mp \frac{k_x^2k_{z2}k_{z3}}{q^2} & \mp \frac{k_xk_yk_{z2}k_{z3}}{q^2} & k_xk_{z3} \\ \mp \frac{k_xk_yk_{z2}k_{z3}}{q^2} & \mp \frac{k_y^2k_{z2}k_{z3}}{q^2} & k_yk_{z3} \\ \mp k_xk_{z2} & \mp k_yk_{z2} & q^2 \end{matrix} \right)\,.
\end{align}
The quantities $T^{(s)}$ and $T^{(p)}$ correspond to the generalized transmission coefficients for the $s$- and $p$-polarized fields respectively that takes into account all internal reflections in the slab. These coefficients are based on the well-known Fresnel reflection and transmission coefficients from medium $i$ to medium $j$,
\begin{align}
    r_{ij}^{(p)}=\frac{k_{zi}\epsilon_j-k_{zj}\epsilon_i}{k_{zi}\epsilon_j+k_{zj}\epsilon_i}\;,\quad r_{ij}^{(s)}=\frac{k_{zi}-k_{zj}}{k_{zi}+k_{zj}}\;,\\
    t^{(p)}_{ij}=\frac{2n_in_jk_{zi}}{k_{zi}\epsilon_j+k_{zj}\epsilon_i}\;,\qquad
    t^{(s)}_{ij}=\frac{2k_{zi}}{k_{zi}+k_{zj}}\,.
\end{align}
Thus, $T^{(s,p)}$ can be calculated as \cite{Sipe,lau2007}
\begin{equation}\label{eq:T21}
    T_{21}(q,z')=\frac{\left[e^{-\ii k_{z2}z'}+r_{23}e^{\ii k_{z2}(2a+z')}\right]}{1-r_{23}r_{21}e^{\ii k_{z2}(2a)}}\times t_{21}\,,
\end{equation}
and
\begin{equation}
    T_{23}(q,z')=\frac{\left[e^{\ii k_{z2}(a+z')}+r_{21}e^{\ii k_{z2}(a-z')}\right]}{1-r_{23}r_{21}e^{\ii k_{z2}(2a)}}\times t_{23}\,,
\end{equation}
where the corresponding Fresnel coefficient for the $s$- and $p$-polarized vectors must be used.

As explained in Sec. (\ref{sec:GFs}), the $s$-polarized field is direction independent, and there is no distinction in directionality for $T^{(s)}$. However, the $p$-polarized field is direction dependant and the coefficient $T^{(p)}$ needs to be treated differently for upward and downward propagating waves. For the $p$-polarized field one can write,
\begin{equation}
    T^{(p)}_{21}(q,z')=T^{(p+)}_{21}(q,z')+T^{(p-)}_{21}(q,z')\,,
\end{equation}
which splits the transmission into an upward field generated in medium 2 that transmits to medium 1, namely $T^{(p+)}_{21}$, and a downward wave generated in medium 2 that also transmits to medium 1, namely $T^{(p-)}_{21}$. From Eq. (\ref{eq:T21}) one can distinguish,
\begin{align}
    T^{(p+)}_{21}(q,z')=\frac{e^{-\ii k_{z2}z'}}{1-r^{(p)}_{23}r^{(p)}_{21}e^{\ii k_{z2}(2a)}}\times t^{(p)}_{21}\,,\\
    T^{(p-)}_{21}(q,z')=\frac{r^{(p)}_{23}e^{\ii k_{z2}(2a+z')}}{1-r^{(p)}_{23}r^{(p)}_{21}e^{\ii k_{z2}(2a)}}\times t^{(p)}_{21}\,.
\end{align}
This distinction becomes important when we want to construct the GF of our system and it was not considered in Ref. \cite{lau2007}. 

For completeness,the same analysis is applied to the GF in Eq. (\ref{eq:gf-}) for waves generated in medium 2 and transmitting to medium 3. Here,
\begin{align}
    T_{23}^{(p+)}(q,z')=\frac{r_{21}^{(p)}e^{\ii k_{z2}(a-z')}}{1-r^{(p)}_{23}r^{(p)}_{21}e^{\ii k_{z2}(2a)}}\times t^{(p)}_{23}\,,\\
    T_{23}^{(p-)}(q,z')=\frac{e^{\ii k_{z2}(a+z')}}{1-r^{(p)}_{23}r^{(p)}_{21}e^{\ii k_{z2}(2a)}}\times t^{(p)}_{23}\,,
\end{align}
where similarly,  $T^{(p+)}_{23}$ describes an upward $p$-wave generated in medium 2 that transmits to medium 3 after multiple reflections and $T^{(p-)}_{23}$ a downward $p$-wave generated in medium 2 that also transmits to medium 3 after multiple reflections.

\section{Pump Field with Multiple Reflections}
\label{ap:pump}

\subsection{Gaussian angular spectrum}
\label{ap:pumpGaus}

\begin{figure}[ht]
\centering
\includegraphics[scale=1]{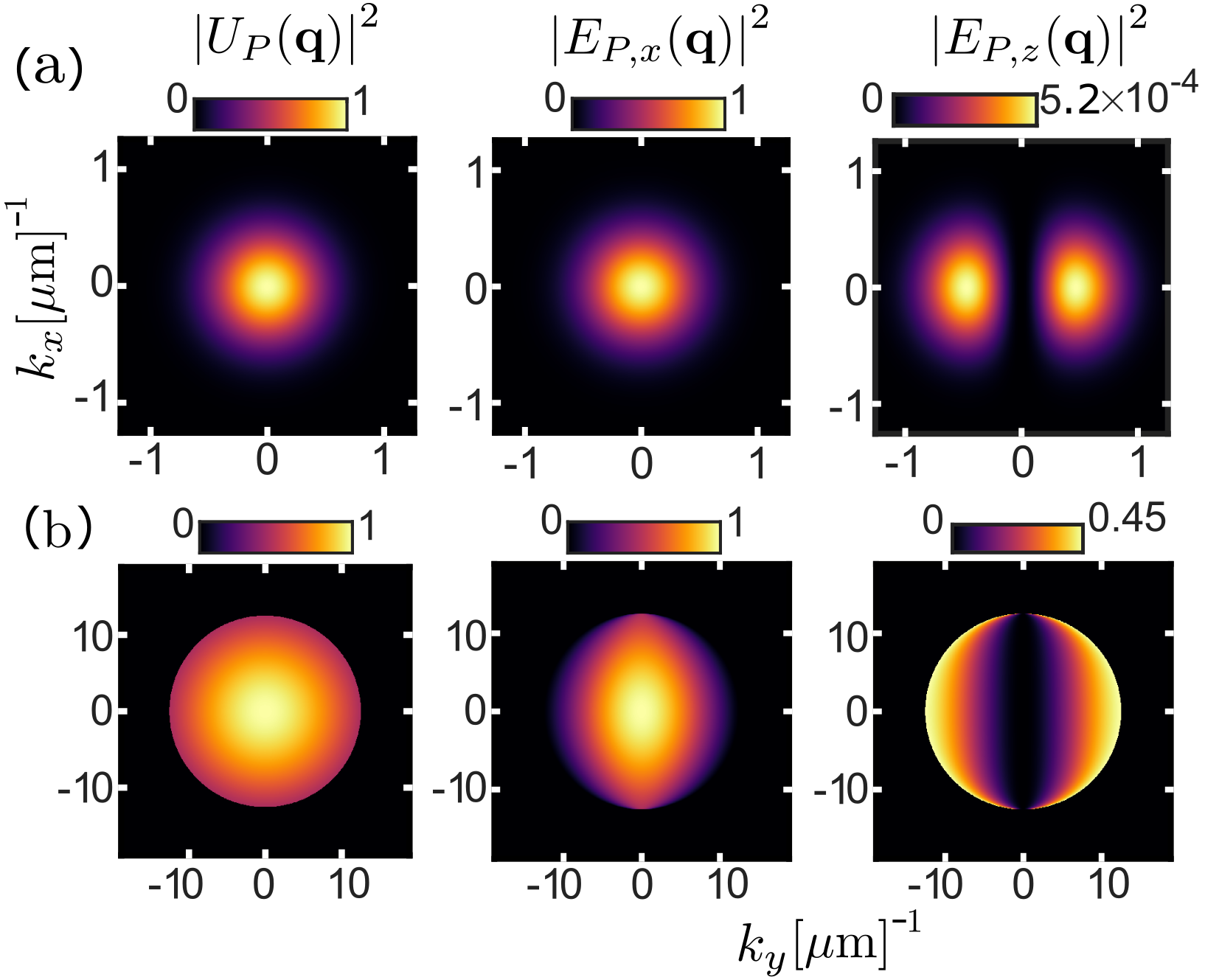}
\caption{Intensity of the spectrum, $x$- and $z$-polarized components of a pump field of wavelength $\lambda_p=500$ nm in (a) for $w=6.6\times10^5$ m$^{-1}$ (paraxial regime) and (b) for a width $w=2\times10^7$ m$^{-1}$ (nonparaxial regime).}
\label{fig:pump}
\end{figure}

To test the contribution of the longitudinal polarization component of the pump beam to the photon-pair generation process we take the Gaussian angular spectrum defined in ($\ref{eq:pump_spec}$) and introduce it to the pump field defined in (\ref{eq:pump}). Using $\lambda_p=500$ nm as the pump wavelength, we can observe in Fig. \ref{fig:pump} the intensity of the spectrum, $x$- and $z$-component of the field in Fourier domain for two spectral pump widths $w$. For a width $w=6.6\times10^5$ m$^{-1}$ Fig. \ref{fig:pump}(a) shows that the spectrum of spatial frequencies are much smaller than the wave-vector ($q<<k_p=2\pi/\lambda_p$), this is the so-called nonparaxial regime (weakly focused beam in position space). Here, $U_p(\mathbf{q})$ corresponds to the angular spectrum of a Gaussian beam of width $W=3\mu$m defined as $u_p(x,y)\propto e^{-(x^2+y^2)/W^2}$. In this regime, the contribution of the longitudinal component of the pump field is very small compared to the transversal component, as it can be seen in Fig.\ref{fig:pump}(a) where the maximum  Max$[E^2_{P,z}(\mathbf{q})]$ is about $\sim10^4$ weaker than Max$[E^2_{P,x}(\mathbf{q})]$. In this regime, the individual plane waves in the superposition (\ref{eq:E_p_fourier}) can be approximated to be only $x$-polarized, due to the small contribution of the $z$-component of the pump electric field. For comparison, we show in Fig. \ref{fig:pump}(b) we show the same quantities for a width $w=2\times10^7$ m$^{-1}$. This is a nonparaxial scenario where the angular spectrum saturates to values close to $\sim k_p$. The Here, the angular spectrum corresponds to a tightly focused beam and the intensity of the $z$-component of the pump is comparable to the $x$-component.


\subsection{Pump field inside the slab}
\label{ap:pumpCalc}

Right before the interface at $z=-a$ we take an $x$-polarized electric pump with field components polarized in the direction of propagation, $\mathbf{E}_p(\mathbf{q},z=-a)=(E_{p,x},0,E_{p,z})$, where its longitudinal component $E_{p,z}$ is calculated from from $E_{p,x}$ using the transverse condition $\mathbf{k}_p\cdot\mathbf{E}_p=0$. This results in
\begin{equation}\label{eq:ExEz}
    E_{p,z}(\mathbf{q},z=-a)=-\frac{k_x}{k_{z,3p}}E_{p,x}(\mathbf{q},z=-a)\,.
\end{equation}

We express the pump field at the interface in terms of the $s$- and $p$-polarized fields as
\begin{align}
\nonumber \mathbf{E}_p&(\mathbf{q},z=-a)\\
\nonumber &={E}_{p,x}(\mathbf{q},z=-a)\hat{x}+{E}_{p,z}(\mathbf{q},z=-a)\hat{z}\\
&={E}^{(s)}_p(\mathbf{q},z=-a)\hat{s}+{E}^{(p)}_p(\mathbf{q},z=-a)\hat{p}_{3+}\,,
\end{align}
where we used the definitions of $\hat{s}$ and $\hat{p}$ in (\ref{eq:s}) and (\ref{eq:p}) for medium 3. After solving we find that
\begin{align}
    E_p^{(s)}(\mathbf{q},z=-a)=\frac{k_{z,3p}k_y}{qk_x}{E}_{p,z}(\mathbf{q},z=-a)\,, \label{eq:Eps}\\
    E_p^{(p)}(\mathbf{q},z=-a)=\frac{k_{3p}}{q}{E}_{p,z}(\mathbf{q},z=-a)\,,   \label{eq:Epp} 
\end{align}
where Eq.~(\ref{eq:ExEz}) still holds. Notice that we have a singularity at $k_x=k_y=0$, which describes a plane wave normal to the interface. For this special point, we take the field to be only $x$-polarized, that is $\mathbf{E}_P(\mathbf{q}=0,z=-a)={E}_{P,x}(\mathbf{q}=0,z=-a)\hat{x}$. 

To describe the transmission of the field to the slab, one can calculate a corresponding coefficient associated to the $s$- and $p$-polarized pump field coming from the outside (medium 3) to the inside of the slab (medium 2), which goes along the lines to the generalized Fresnel transmission coefficients (Eq. (\ref{eq:T21})). The calculation of the coefficients is done in the appendix \ref{ap:Q} and take the form
\begin{equation}\label{eq:Q}
    Q_{32}(q,z)=\frac{e^{\ii k_{z,2p}(a+z)}(1+r_{21}e^{-\ii k_{z,2p}(2z)})}{1-r_{23}r_{21}e^{\ii k_{z,2p}(2a)}}\times t_{32}\,,
\end{equation}
where the first term
\begin{equation}
    Q^{(p+)}_{32}(q,z)=\frac{e^{\ii k_{z,2p}(a+z)}}{1-r_{23}r_{21}e^{\ii k_{z,2p}(2a)}}\times t_{32}\,,
\end{equation}
describes upward waves and the second term 
\begin{equation}
    Q^{(p-)}_{32}(q,z)=\frac{r_{21}e^{\ii k_{z,2p}(a-z)}}{1-r_{23}r_{21}e^{\ii k_{z,2p}(2a)}}\times t_{32}\,,
\end{equation}
downward waves inside the slab. This distinction is again only important for the $p$-polarized fields, since a $p$-field changes direction when the field propagates upward and downward (see Eq. (\ref{eq:p})), while the $s$-field remains unchanged. Notice that the factors are $z$-dependent and consider the multiple reflections of the pump field in the slab. 

Thus, we arrive at Eq. (\ref{eq:Epump}), where the pump field inside the slab is
$    \mathbf{E}_p(x,y,z)=\frac{1}{(2\pi)^2}\int\dd\mathbf{q}\;\mathbf{E}_p(\mathbf{q},z)\,e^{\ii \mathbf{q\cdot r}_\perp}\,,
$
with
\begin{align}
    \nonumber &\mathbf{E}_p(\mathbf{q},z)=Q^{(s)}(q,z)E^{(s)}_p(\mathbf{q},z=-a)\hat{s}\\
    &+E^{(p)}_p(\mathbf{q},z=-a)\left(Q^{(p+)}(q,z)\hat{p}_{2+}+Q^{(p-)}(q,z)\hat{p}_{2-}\right)\,,
\end{align}
where again the $s$- and $p$- directions for each plane wave in the expansion are defined as in Eqs. (\ref{eq:s},\ref{eq:p}) for medium 2. Note that,  for a plane inside the slab located at $z\in\{-a,0\}$, the term $Q^{(s)}(q,z)E^{(s)}_p(\mathbf{q},z=-a)\hat{s}$ in equation (\ref{eq:Epump}) contains the sum of all the direction-independent $s$-polarized fields, while the direction-dependent term $E^{(p)}_p(\mathbf{q},z=-a)\left(Q^{(p+)}(q,z)\hat{p}_{2+}+Q^{(p-)}(q,z)\hat{p}_{2-}\right)$ takes all $p_+$ (upward) and $p_-$ (downward) fields that add up all after transmission to the slab and multiple reflections.  

Finally, using the definitions in Eqs. (\ref{eq:s},\ref{eq:p}), we can insert the computed Eqs. (\ref{eq:Eps}) and (\ref{eq:Epp}) into Eq. (\ref{eq:Epump}) to express the pump field inside the slab in cartesian coordinates.

\subsection{Generalized transmission coefficient for the pump field inside the slab}
\label{ap:Q}

A field coming from region 3 transmits into region 2 at the interface $z=-a$, then it propagates to $z$ gaining a phase factor of $\alpha=e^{ik_z(a-\abs{z})}$. The field then propagates to $z=0$ and reflects back to $z$ gaining an extra phase of $\beta=e^{ik_z\abs{z}}r_{21}e^{ik_z\abs{z}}$. The filed now propagates to $z=-a$ and reflects again to $z$ with an extra phase of $\gamma=e^{ik_z(a-\abs{z})}r_{23}e^{ik_z(a-\abs{z})}$. Finally this process is repeated indefinitely due to multiple reflections. The field inside the slab at $z$ can then be describe through the recurrence
\begin{align}
    \nonumber Q^{(s,p)}(q,z)=(\alpha+\beta\alpha+\gamma\beta\alpha+\gamma\beta^2\alpha+\gamma^2\beta^2\alpha
    \\+\gamma^2\beta^3\alpha+\gamma^3\beta^3\alpha+...)t_{32}\,,
\end{align}
which converges to
\begin{equation}
   Q^{(s,p)}(q,z)=\frac{\alpha(1+\beta)}{1-\beta\gamma}t_{32}\,,
\end{equation}
leading to Eq.~(\ref{eq:Q}). If we split the recurrence into the upward and downward waves at $z$ we find $Q^{(p+)}(q,z)=\frac{\alpha}{1-\beta\gamma}t_{32}\,,$ and $Q^{(p-)}(q,z)=\frac{\alpha\beta}{1-\beta\gamma}t_{32}\,,$

\section{Quantum Polarization Tomography}
\label{ap:QPT}

Following Ref. \cite{James}, the classical Stokes parameters can be generalized to measure the state of multiple photon beams. We use the $\{\ket{HH},\ket{HV},\ket{VH},\ket{VV}\}$ basis to express the polarization density matrix $\hat{\rho}$ which is reconstructed by 16 projections into a set of states $\{\ket{\psi_\nu}\}_{\nu \in [\![1;16]\!]}$. We employ the same basis states $\ket{\psi_\nu}$ as in Ref.~\cite{James}. The orientation of the orthogonal basis vectors $\ket{H}$ and $\ket{V}$ can in principle be chosen freely, however will influence the shape of the density matrix (but importantly not the degree of entanglement, which is independent of the basis). We choose here a convention for $\ket{H}$ and $\ket{V}$ that relates to an experimental scenario, where photon pairs are first collimated by a lens and then projected into different measurement bases by polarizers fixed in the laboratory coordinate frame $\hat{x}=\ket{H}$, $\hat{y}=\ket{V}$. Following this, we define a new set of vectors $\{\hat{x}',\hat{y}'\}$ in the plane $\{\hat{\theta},\hat{\varphi}\}$, which for a propagation direction $\varphi$ follow from

\begin{equation}
     \begin{cases}
    \hat{x}'&=\cos{\varphi}\, \hat{\theta}-\sin{\varphi}\, \hat{\varphi} \\
    \hat{y}'&=\sin{\varphi}\, \hat{\theta}+\cos{\varphi}\, \hat{\varphi}
         \end{cases}.
\end{equation}
When collected and collimated as in an experimental scheme, these vectors coincide with the lab coordinate frame and therefore $\hat{x}'=\ket{H}$, $\hat{y}'=\ket{V}$.


The Schmidt entanglement parameter K (Schmidt number) quantifying the degree of entanglement of the pair is given by \cite{grobe,eberly}:
\begin{equation}
    K=\frac{1}{Tr\left(\hat{\rho}_i^2\right)}=\frac{1}{Tr\left(\hat{\rho}_s^2\right)}\,,
\end{equation}
where $\hat{\rho}_s$ and $\hat{\rho}_i$ are the reduced density matrices corresponding to the single particle wavefunctions which are
obtained from the partial trace of the biphoton density matrix: $\hat{\rho}_{s,i}=Tr_{s,i}(\hat{\rho})$.

To investigate the effect of a finite collection angle on degree of polarization entanglement, we perform the following calculation. We assume an objective/lens with numerical aperture NA is collecting all signal and idler photons going in any pairs of emission angles within the cone of angle $\theta=\arcsin(\mathrm{NA})$. To perform a numerical analysis corresponding to this scenario, we first integrate over the detection probabilities of pairs coming from all the angles in that collection cone for a fixed projection measurement. The integral is performed over the solid angles $\dd\Omega = \sin\theta \dd\theta\dd\varphi$ for both the signal and idler photons. We repeat this to find the 16 quantities for the projective measurements, that now allows us to reconstruct the polarization density matrix for the total collected pairs. Finally, the purity of the state is calculated through $P = Tr(\hat{\rho}^2)$.


\bibliographystyle{ieeetr.bst}
\bibliography{Refs.bib}

\end{document}